\documentclass[a4paper,fleqn,usenatbib]{mnras}

\usepackage{mathptmx}
\usepackage{txfonts}

\usepackage[T1]{fontenc}
\usepackage{ae,aecompl}

\usepackage{graphicx}	
\usepackage{amssymb}	
\usepackage{longtable}
\usepackage{pdflscape}
\usepackage{natbib}
\usepackage{textcomp}
\usepackage{ulem}

\title[Rotation and spots in normal A and Am/Fm stars]{Rotation and spots in normal A and Am/Fm stars}

 \author[O. Trust et al.]{
Otto Trust$^{1}$\thanks{E-mail: otrust@must.ac.ug},
Edward Jurua$^{1}$\thanks{E-mail: ejurua@must.ac.ug},
Peter De Cat$^{2}$
  and 
Santosh Joshi$^{3}$
 \\
 $^{1}$Department of Physics, Mbarara University of Science and Technology, P.O. Box 1410, Mbarara, Uganda\\
$^{2}$Royal Observatory of Belgium, Ringlaan 3, B-1180 Brussel, Belgium\\
$^{3}$Aryabhatta Research Institute of Observational Sciences, Manora Peak, Nainital- 263002, India
 }

\date{Accepted XXX. Received YYY; in original form ZZZ}

\pubyear{2019}

\begin{document}
\label{firstpage}
\pagerange{\pageref{firstpage}--\pageref{lastpage}}
\maketitle

\begin{abstract}

  Frequency analysis of long-term ultra-precise photometry can lead to precise values of rotation frequencies of rotating stars with ``hump and spike'' features in their periodograms. Using these features, we computed the rotation frequencies and amplitudes. The corresponding equatorial rotational velocity ($\rm V_{rot}$) and spot size were estimated.  On fitting the autocorrelation  functions of the light-curves with the appropriate model, we determined the starspot decay-time scale. The $\rm V_{rot}$ agrees well with the projected rotational velocity ($\rm \nu~sin$~$i$) in the literature.  Considering a single circular and black spot we estimate its radius from the amplitude of the ``spike''. No evidence for a significant difference in the average ``spike'' amplitude and spot radius was found for Am/Fm and normal A stars. Indeed, we derived an average value of $\rm \sim 21\pm2$ and $\rm \sim 19\pm2$\,ppm for the photometric amplitude  and of $\rm 1.01\,\pm\,0.13$ and $\rm 1.16\,\pm\,0.12$\,$\rm R_E$ for the spot radius (where  $\rm R_E$ is the Earth radius), respectively. We do find a significant difference for the average spot decay-time scale, which amounts to $3.6\pm0.2$ and $1.5\pm0.2$ days for Am/Fm and normal A stars, respectively. In general, spots on normal A stars are similar in size to those on Am/Fm stars, and both are weaker than previously estimated. The existence of the ``spikes'' in the frequency spectra may not be strongly dependent on the appearance of starspots on the stellar surface. In comparison with G, K and M stars, spots in normal A and Am/Fm stars are weak which may indicate the presence of a weak magnetic field.

\end{abstract}




\section{Introduction}
 The main sequence A-type stars are intermediate-mass stars which show various interesting phenomena such as chemical peculiarities, pulsation, activity and rotation. The chemically peculiar (CP) stars \citep{1974ARA&A..12..257Pp, 1988A&AS...76..127R, Monier_2019} can be distinguished from the normal A stars \citep{1986A&AS...64..173aa, 2004IAUS..224....1A} based on the enhancement of the absorption lines in their spectra. The optical spectra of the CP stars are characterized by strong lines of silicon, metals and/or rare-earth elements and weak lines of calcium. The CP stars are dominated by metallic line (Am/Fm) stars and are sub-divided into two main groups: the Am/Fm stars, (CP1; having an over-abundance of iron-group metals and an under-abundance of He, Ca and/or Sc in their atmospheres), and the magnetic Ap stars (CP2; showing photometric and magnetic variability and having abnormally strong lines of Si, Cr, Sr and rare-earth  elements in their spectra) \citep{1970PASP...82..781C, 1974ARA&A..12..257Pp, 2009A&A...503..945F}.

 The atomic diffusion is thought to be the dominant process responsible for the chemical anomalies observed in the CP stars \citep{1970ApJ...160..641M,1970ApJ...162L..45W,khokh, hui, 2003ASPC..305..199T, 2011sf2a.conf..253T}. Due to radiation pressure and gravitational settling, chemical elements undergo an upward/downward drift into/out of the atmosphere. Depending on the ionization state and the atomic properties of the elements and in the absence of mixing, an equilibrium is attained. At equilibrium, the elements are vertically stratified \citep{2002A&A...384..545R, 2015AASP....5....3K} and the abnormal chemical abundances manifest in the atmospheres. In Am/Fm stars, when the magnetic fields are detected, they are ultra-weak and relatively uniform \citep[e.g;][]{2010A&A...523A..40A, blaz2, blaz, 2017MNRAS.468L..46N,2019ApJ...883..106C} which produce evenly distributed abundances in the atmospheres. On the other hand, Ap stars possess strong global magnetic fields of the range $\rm 200~G-30~kG$ \citep{2007pms..conf...89P, 2014IAUS..302..255B}. In these stars, depending on the magnetic field strength which may suppress convection and the inclination of the field lines, patches of different abundances are produced on the surface \citep{1970ApJ...160..641M, 1973IrAJ...11...32M}.

 The atomic diffusion can be counteracted by mixing. The currently known mixing processes include meridional circulation caused by rotation, and convection. This implies that the CP stars should be rotating more slowly relative to normal A stars. The majority of Am/Fm stars are indeed slow rotators  \citep{2008JKAS...41...83T,2008A&A...483..891F, stateva}, with $\rm \nu~sin$~$i < 120~\rm km~s^{-1}$  \citep{Abt2009}. The slow rotation in Am/Fm stars is attributed to the tidal synchronisation in the case of close binaries \citep{khokh, kun, balona11}. The Am/Fm stars are often components of binaries \citep{1961ApJS....6...37A, 1976A&A....50..435V, 1998CoSka..27..179N, 2000IAUS..200P.161D, stateva} with orbital periods between 1 and 10 days \citep{2014A&A...564A..69S}. This makes rotation one of the most important phenomena in stellar physics. The availability of accurate and precise values of rotation periods provides an important platform in understanding other stellar phenomena  such as pulsation \citep{1988AcA....38...61D,1998A&A...334..911S} and overshooting \citep{2004ApJ...601..512B, 2019MNRAS.485.4641C} and allows to draw conclusions about the processes that may  be linked with the observed chemical abundances in stellar atmospheres of Am/Fm stars \citep{2001astro.ph.11179T, 2014PhDT.......131M}. The other reported success of the rotationally induced mixing is the ability to explain the origin of the Boesgaard lithium gap \citep{1986ApJ...303..724B} in F stars \citep{2000ApJ...544..944S}. Apart from using spectroscopy, where the rotation rate is retrieved from the broadening of the absorption lines \citep{1989ssis.book.....K}, stellar rotation period can be determined from seismic splitting \citep{1981ApJ...244..299S,1981Natur.293..443C}, spot induced rotational modulation \citep{1947PASP...59..261K, 2009A&A...506..245M} and chromospheric activity especially Ca\,II emission \citep{1984ApJ...279..763N}.

 Thanks to the availability of high-precision and almost continuous photometric data from space missions like the nominal {\it Kepler} mission \citep{borucki10}, K2 \citep{2014PASP..126..398H} and the Transiting Exoplanet Survey Satellite \citep[{\it TESS};][]{1.JATIS.1.1.014003}, from which precise rotation periods can be obtained. \citet{balona13} derived rotation periods of a large number of A-type stars from the original {\it Kepler} field \citep{borucki10} based on spot induced rotational modulation. Starspots and flares are claimed to have been discovered in rotationally modulated normal A and Am/Fm stars \citep{2012MNRAS.423.3420B, balona13, balona15, balona16}. \citet{balona13} obtained a mean rotational modulation amplitude, interpreted as a reflection of the starspot size, of 875 normal A stars as 541\,$\pm$\,86~ppm and \citet{balona15} found 93\,$\pm$\,25~ppm for 10 Am/Fm stars. As starspots and flares are signatures of a high magnetic field, this would imply a strong magnetic field in these stars.  \citet{2017MNRAS.466.3060P} confirmed the presence of flare-like structures in only some stars studied by \citet{2012MNRAS.423.3420B,balona13}. However, after consideration of different possibilities, \citet{2017MNRAS.466.3060P} found that the flare-like structures have a high possibility of originating from cool and unresolved companions.

 The discovery of spot based rotational modulation in normal A and Am/Fm stars was unexpected. They possess a thin sub-surface convective envelope, which increases in depth with decreasing effective temperature and decreasing surface gravity \citep{1984ApJ...286L..19G, Abt2009}. This implies that they lack strong tangled magnetic fields which would be created by the dynamo effect resulting from the convective behavior of the material with rotation \citep{Branden2000} like in G, K and M stars.  \cite{2007A&A...475.1053A} reported that stellar magnetic fields of magnitude below 300\,G are not detected. However, ultra-weak global magnetic fields have been observed in a number of stars, e.g; normal A-type stars, like Vega \citep[0.6\,$\pm$\,0.3\,G;][]{2009A&A...500L..41L}, \citep[1.0\,G and 1.4\,G;][]{2010A&A...523A..41P}, \citep[7.0\,G;][]{2014A&A...568C...2P} and Am/Fm stars, like  Sirius A ($\rm 0.2\,\pm\,0.1\,G$) \citep{2011A&A...532L..13P}, Alhena ($\rm around\, 5\,\pm\,3\,G$) \citep{blaz}, $\beta$\,UMa ($\rm 1\,\pm\,0.8\,G$) and $\theta$\,Leo ($\rm 0.4\,\pm\,0.3\,G$) \citep{2016A&A...586A..97B} and $\rm\rho~Pup$ ($\rm 0.29\,\pm\,0.32\,G$) \citep{2017MNRAS.468L..46N}. \cite{2019ApJ...883..106C} theoretically predicted that A-stars like Vega, Sirius, $\beta$\,UMa, $\theta$\,Leo and Alhena can have magnetic fields of magnitude of 1\,--\,10\,G generated by dynamo action in the thin He\,II convective subsurface zone and transported to the surface by magnetic bouyancy.  Magnetic fields with such magnitude and the thin convective zone could support, if any, weak starspots \citep{blaz, 2017MNRAS.468L..46N, 2019ApJ...883..106C}. Therefore, spot sizes in normal A and Am/Fm stars given by \citet{balona13} and \citet{balona15}, respectively, could be overestimated.
 
\begin{figure}
  \includegraphics[width=\columnwidth]{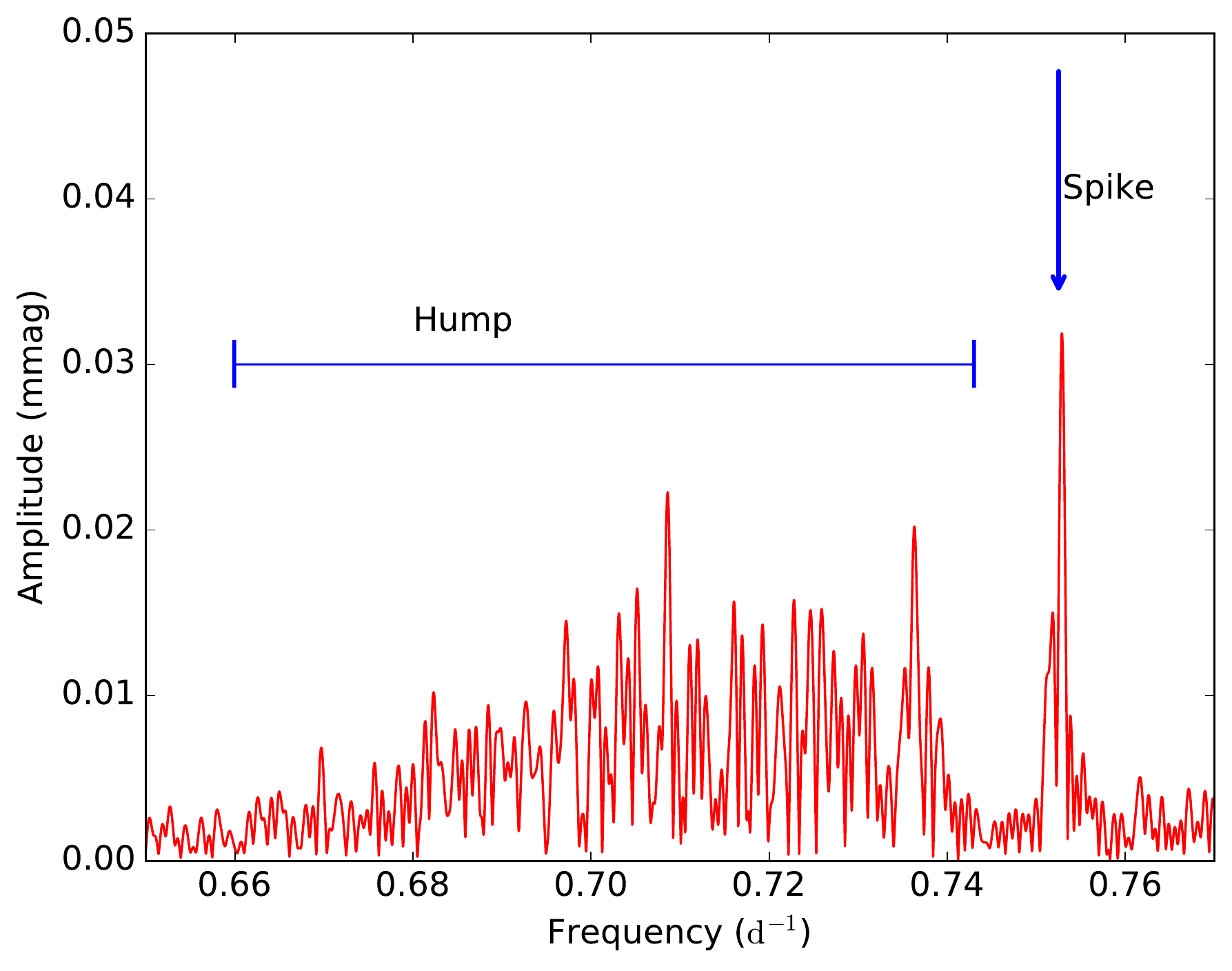}
  \caption{An amplitude spectrum of KIC\,6192566 showing a typical hump and a spike. The spike is not strictly sharp.}
  \label{spike}
 \end{figure}

 In addition, \citet{balona13} observed in the frequency spectra of 135 stars of the 875 normal A stars a sharp peak (``spike'') on the high frequency side of a broad hump of very close frequencies (``hump'') as shown in Fig.~\ref{spike}. The sharp frequency and the broad hump were seen to have harmonics of lower amplitudes in most cases. \citet{balona15} observed the same scenario in some Am/Fm stars. \citet{2014MNRAS.441.3543B} suspected the sharp peak corresponds to the rotational frequency and its amplitude to be related to the starspot size, but the hump remained unexplained. Recently, \citet{2018MNRAS.474.2774S} named these stars "hump and spike" stars. \citet{2018MNRAS.474.2774S} carefully compared the visibility curves for the global Rossby waves ({\it r}\,modes) \citep{1978MNRAS.182..423P} with the frequency spectra of {\it Kepler} light curves and concluded that the broad humps as observed for these stars are induced by {\it r}\,modes and the spike structures are the rotation frequencies induced by one or more spots. 
 
 To search and study the pulsational variabilities in Ap and Am/Fm stars, a dedicated ground-based project the ``Nainital-Cape Survey'' was initiated between astronomers of India and South Africa. However, with time, astronomers from other institutions in other countries joined this programme making it a multi-national collaborative project and a number of results are published \citep[e.g;][]{2000BASI...28..251A, 2001AA...371.1048M,2003MNRAS.344..431J,2006A&A...455..303J, 2009A&A...507.1763J, 2010MNRAS.401.1299J, 2012MNRAS.424.2002J, 2016A&A...590A.116J, 2017MNRAS.467..633J}. The success of the ``Nainital-Cape Survey'' and the discovery of ``hump and spike'' features in the frequency spectra \citep{balona13,balona15,2018MNRAS.474.2774S} of a number of normal A and Am stars  motivated us to initiate another tri-national collaborative group among the astronomers from India, Uganda and Belgium. This paper is the first of the series involving the ``hump and spike'' stars aiming to present rotation frequencies and velocities, starspot sizes and starspot decay life-times based on the ``hump and spike'' features. Only normal A and Am/Fm stars in the original {\it Kepler} field with ``hump and spike'' features in their frequency spectra are investigated. This paper is organized follows.  The adopted target selection criteria and the data reduction are described in Section~\ref{obs}. The results and their discussion are presented in Section~\ref{res_dis}. The conclusion drawn from our study is given in Section~\ref{concl}.

\section{Target Selection and Data Reduction}
\label{obs}

\subsection{Selection criteria}

The samples in this study were drawn from stars analysed by \citet{balona13}, \citet{balona15}, and \citet{2016AJ....151...13G}. Of the 875 rotationally modulated main sequence normal A stars studied by \citet{balona13} and the 15 Am/Fm stars from the nominal {\it Kepler} mission analysed by \citet{balona15}, only the normal A and Am/Fm stars with ``hump and spike'' features in their frequency spectra were considered. Additional Am/Fm stars with ``hump and spike'' features in the frequency spectra were derived from the sources analysed by \citet{2016AJ....151...13G}  under the LAMOST-{\it Kepler} project \citep{2015ApJS..220...19D}. A total of 170 ``hump and spike'' stars were obtained of which 131 are part of the normal A stars reported by \citet{balona13}, 3 are within the sample studied by \citet{balona15}, and 36 are from \citet{2016AJ....151...13G}.

\subsection{Data reduction}
\label{space}
For this study, we used the high precision and almost continuous photometric data obtained by the {\it Kepler} mission. The data are available on the Barbara A. Mikulski Archive for Space Telescopes (MAST) website\footnote{http://archive.stsci.edu/}. The stars in the sample were observed in the long-cadence mode with an observation every 29.4 minutes. A few of the ``hump and spike'' stars were also observed in short cadence with a time interval of almost 1 minute. Since we are interested in low frequencies, long-cadence data are sufficient to capture long period variations. The characteristics of long-cadence data are discussed by \citet{2010ApJ...713L.120J}. Within its orbit of 372.5 d, the {\it Kepler} space craft performs a $\rm 90^{\circ}$ roll every 3 months to keep its solar panels aligned with the sun. This resulted into quarters named $\rm Q_0 ~upto~Q_{17}$. The commissioning run ($\rm Q_{0}$) only lasted for 10\,days, quarters $\rm Q_{1}$ and $\rm Q_{17}$ were only 1 month long. This study uses the data for all the quarters, unlike in \citet{balona13} where only the data from quarters $\rm Q_0$ to $\rm Q_{12}$ were available. 

The light-curves are stored in the form of a Flexible Image Transport system (FITs) product, which contain tables with a combination of header key words \citep{borucki10, koch}. Among others, the light-curves contain simple aperture photometry (SAP) and pre-search data conditioning (PDC) flux \citep{2016ksci.rept....9T}. The SAP flux results from the application of the {\it Kepler} data processing pipeline, which only uses a basic calibration, and is affected by errors. The dominant sources of errors include: differential velocity aberration and thermal transients from the reaction wheel desaturation and  quarterly rolls \citep{2012PASP..124.1000S, 2014PASP..126..100S}. The PDC module of the data processing pipeline tries to identify and minimise such signal distortions and noise without jeopardising astrophysical signals \citep{2012PASP..124.1000S}. The most recent version of the pipeline is the multi-scale Maximum A Prior (msMAP) pipeline, and this is the one available through MAST. A more detailed description of the characteristics of PDC flux is found in \citet{2012PASP..124..985S} and \citet{2012PASP..124.1000S}. For our purpose, we adopted the PDC flux. The required input for this procedure (barycentric Julian date relative to JD~2\,450\,000.00 and the PDC flux) were extracted from the FITs files. The time was converted to barycentric Julian date relative to JD~2\,454\,950.00 while the PDC flux was converted to millimagnitudes. In different quarters, amplitude variations were observed. To obtain an average amplitude, all the data for all the quarters were stringed together to produce a single light-curve. The detailed analysis of the light-curves and the corresponding results are discussed in Section~\ref{res_dis}.

 \section{Results and Discussion}
\label{res_dis}
\subsection{Rotation frequencies}
\label{period}
\begin{figure}
  \centering
 \includegraphics[width=\columnwidth]{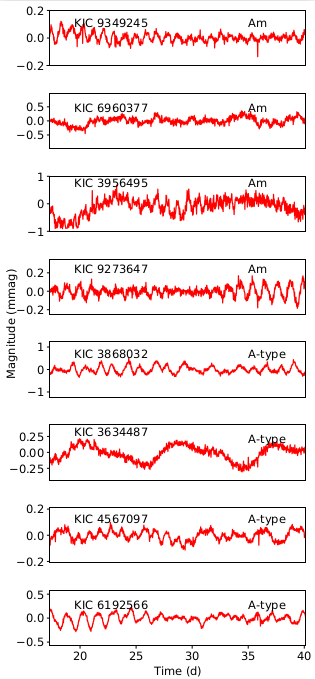}
\caption{Typical light-curves of Am/Fm and normal A stars in the sample of selected stars.}
\label{fig2}
\end{figure}

A precise estimate of the rotation frequency of a star results in a reliable rotational velocity provided that the radius is known. Fig.~\ref{fig2} represents some of the {\it Kepler} light-curves that we analysed. These light-curves were used to search for amplitude variations. Inspecting the light-curves, they seem to be modulated. The modulation mostly comes from the beating of the plethora of {\it r}\,modes. To a lesser extent in such stars, the amplitude modulation could be the result of two or more sinusoids of slightly different frequencies. In addition, we may also not rule out the possibility of having cool backgrounds or close companions responsible for the rotational modulation, though this is expected to be a rare occurrence in our sample. The cooler $\rm T_{eff}$  and greater distance of the background stars would provide weaker total light signal than the study sample. The weaker total light signal would create very small rotational modulation effects which would not always manifest as a measurable effect in our star’s light curve. This implies that the observed signal is intrinsic to the star. The majority of the light-curves display low amplitudes which hampers the visibility of the amplitude variations.  An appropriate tool to display and quantify amplitude variability is the discrete Fourier transform for the unequally spaced data \citep{1982AJ.....87.1608S}. 

\begin{figure*}
  \centering 
 \includegraphics[width=0.8\textwidth]{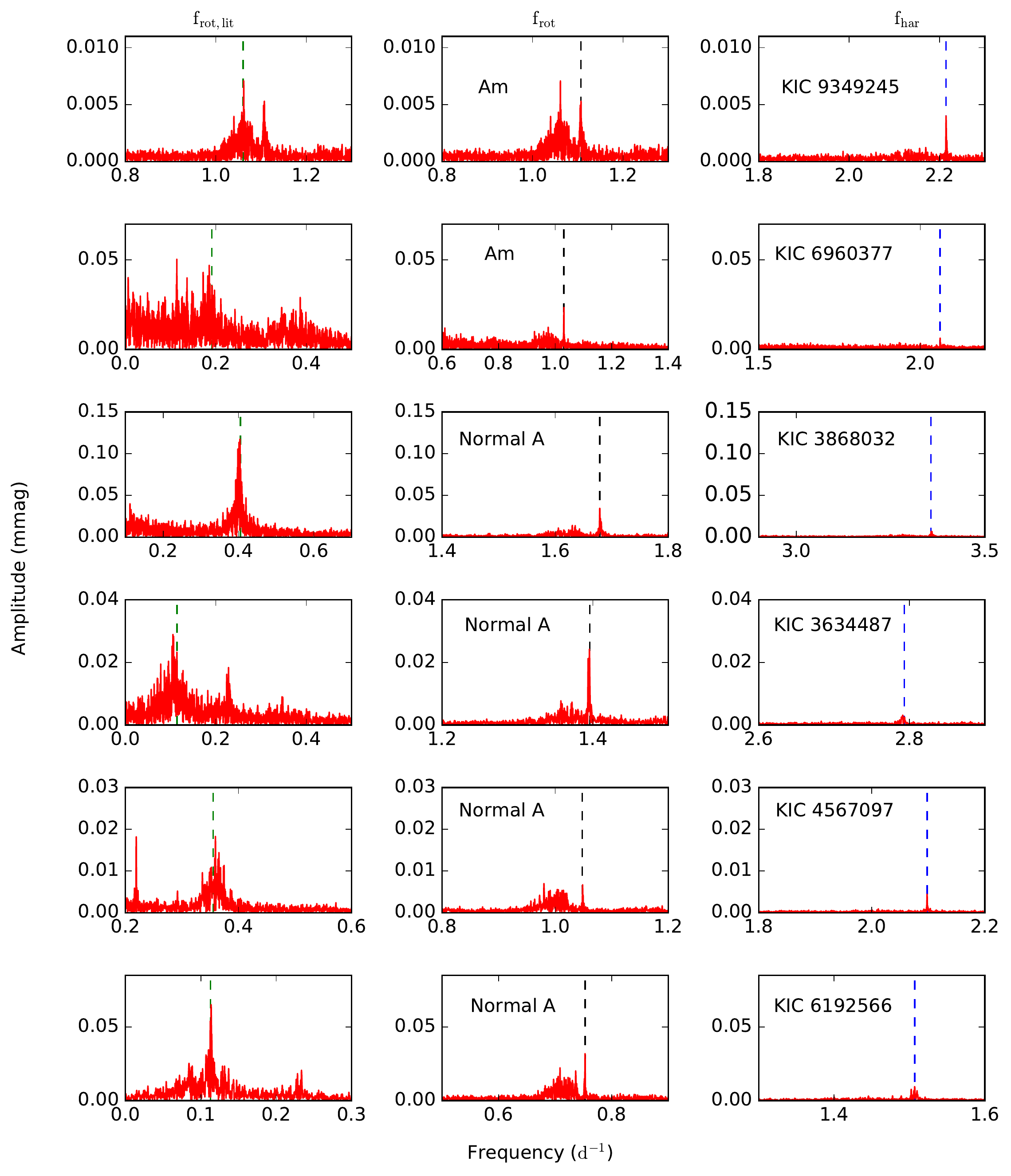}
\caption{Frequency spectra of selected Am/Fm and normal A stars based on the 4-year {\it Kepler} data. The green dashed lines in the panels on the left represent the rotation frequencies ($\rm f_{rot,lit}$) from the literature \citep{balona13,balona15}.  Panels in the middle indicate the ``humps and spikes'' features and those on the right show their first harmonics. The black dashed lines represent the estimated rotation frequencies ($\rm f_{rot}$) and the blue dashed lines represent the frequency equivalent to $\rm 2f_{rot}$.  In the cases where $\rm f_{rot}$ is different from $\rm f_{rot,lit}$, the spike is not the dominant frequency, a characteristic on which $\rm f_{rot,lit}$ was based.}
\label{fig3}
\end{figure*}

For all the stars in our sample, frequencies were calculated from the light-curves using the discrete Fourier fitting technique incorporated in the software package {\it Period04} \citep{2005CoAst.146...53L}.   In Fig.~\ref{fig3}, we present typical periodograms for selected stars, whose light-curves are shown in Fig.~\ref{fig2}.  By entirely visual inspection of the periodograms, we identified the frequencies. The features of much interest in the frequency spectra are broad humps of unresolved frequencies coupled with spikes at their slightly higher frequency end. The ``hump and spike'' features were found in frequency spectra of 131  normal A and 39 Am/Fm stars. Since it is hard to reliably distinguish between true signals and noise at very low frequencies, the part of the periodogram below $\rm 0.05~d^{-1}$ was always ignored. In Fig.~\ref{fig3}, the panels on the left show the rotation frequencies ($\rm f_{rot,lit}$) from the literature \citep{balona13,balona15}. The  $\rm f_{rot,lit}$ values correspond to the dominant frequencies in a given periodogram. The middle panels show ``hump and spike'' features.  Features in the right panels correspond to their harmonics. However, in some stars such harmonics are not visible enough, especially for the humps. The presence of harmonics is a characteristic of stellar rotation  and the presence of spots.
Since the light curve is more of non-sinusoidal shape, the spots create frequency harmonics in the Fourier transform depending on the spot latitude and the inclination angle \citep{2018Ap&SS.363..260C}. 
 
\begin{figure}
  \centering 
 \includegraphics[width=\columnwidth]{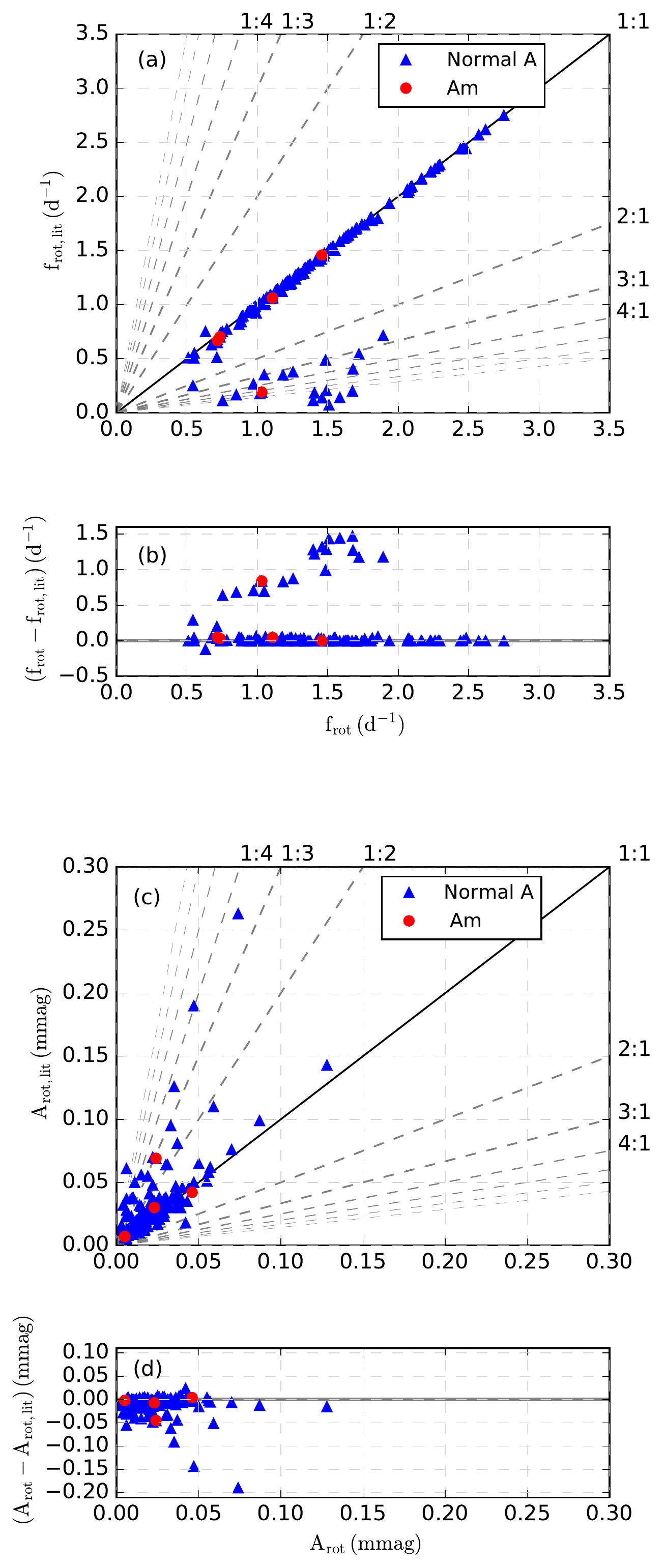}
\caption{In panel (a), comparison of rotation frequencies ($\rm f_{rot}$) with values from previous studies ($\rm f_{rot,lit}$).  Panel (b) shows the residuals in the rotation frequencies. In panel (c),  photometric amplitudes ($\rm A_{rot}$) compared with the values ($\rm A_{rot,lit}$) from the literature \citep{balona13, balona15}.  Panel (d) represents the residuals in the  photometric amplitudes. The full black line is the 1:1 relation, and the dashed lines correspond to harmonics (1:N) and subharmonics (N:1).  In panels (a) and (c), the stars that fall off the 1:1 line have rotation frequencies and amplitudes, respectively, which we have completely corrected.}
\label{fig3b}
\end{figure}

From the periodograms shown in Fig.~\ref{fig3}, spike frequencies were obtained and considered as rotation frequencies \citep{2018MNRAS.474.2774S}.  The results for normal A and Am/Fm stars are depicted in Tables~\ref{table1} and \ref{table_Am}, respectively. The uncertainties in rotation frequencies were computed using the least squares algorithm and found to be of the order of $\rm 10^{-5}~d^{-1}$.
In Fig.~\ref{fig3b}, we compare our rotation frequencies ($\rm f_{rot}$) and photometric amplitudes ($\rm A_{rot}$) with values from previous studies, $\rm f_{rot,lit}$ and $\rm A_{rot,lit}$, respectively, \citep{balona13} and \citep{balona15}. The full black line is the 1:1 relation, and the dashed lines correspond to harmonics (1:N) and subharmonics (N:1). For the stars that lie along the line with 1:1 relation, the spike corresponds with the frequency of the highest amplitude in a given periodogram. In such cases, our results are in good agreement with the previous values. For the stars where this is not the case, the corresponding values of rotation frequency and amplitude are different, and the discrepancy increases with increasing N. The residuals in rotation frequencies and amplitudes from the 1:1 lines are represented in panels (b) and (d), respectively. The stars whose rotation frequencies and amplitudes we have entirely corrected lie off the 1:1 line in the panels (a) and (c).

\subsection{Rotational velocities}
\label{vel}
\begin{figure}
 \centering
 \includegraphics[width=\columnwidth]{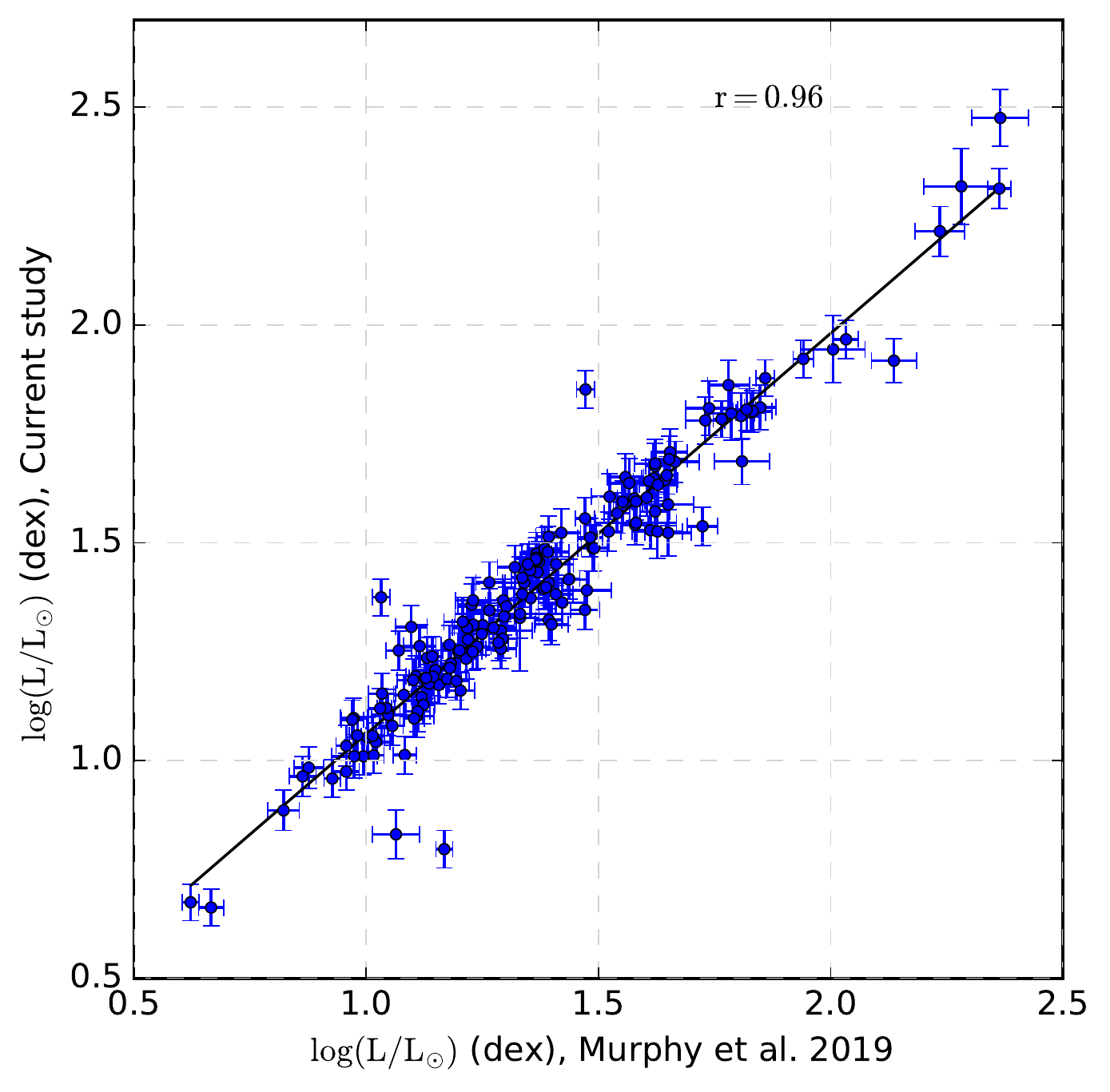}
 \caption{ The $\rm log\left(L/L_{\odot}\right)$ determined by us compared with that computed by \citet{2019MNRAS.485.2380M}. The black solid line corresponds to the best fit. r-value the correlation coefficient is given in the top right corner.}
 \label{fig:LM}
\end{figure}

 Using the resulting rotation frequencies and stellar radii, rotational velocities were calculated from,  \begin{equation}
 \rm V_{rot}=2\pi R_{eq} f_{rot},
 \label{velo}
\end{equation} where $\rm V_{rot}$ is the equatorial rotational velocity in $\rm km~s^{-1}$, $\rm R_{eq}$  is the equatorial stellar radius in km and $\rm f_{rot}$ is the rotation frequency in Hz. We used the mean stellar radius (R) which we calculated from,
\begin{equation}
  {\rm log\left(\frac{R}{R_{\odot}}\right)=0.5log\left(\frac{L}{L_{\odot}}\right)}-2.0{\rm log}\left(T_{\rm eff}\right)+7.52340473,
\end{equation} instead of the $\rm R_{eq}$. In most cases, $\rm R_{eq}$ is identical to R and the difference does not change the $3\sigma$ significance level of the result. We derived the luminosity parameter ($\rm log\left(L/L_{\odot}\right)$) from GAIA parallaxes \citep{2018yCat.1345....0G}, reddening from a 3D model \citep{bayestar, 2019arXiv190502734G} and adopted the effective temperature ($\rm T_{eff}$) from the revised catalog of {\it Kepler} targets for the $\rm Q_{1-17}$ by \citet{2017ApJS..229...30M}.   For stars whose parallaxes are unknown, we considered the radius from the revised catalog of {\it Kepler} targets by \citet{2017ApJS..229...30M}. The resulting values for luminosity, stellar radius and rotational velocities for normal A and Am/Fm stars are listed in Tables~\ref{table1} and \ref{table_Am}. In Fig.~\ref{fig:LM}, our derived $\rm log\left(L/L_{\odot}\right)$ is compared with that determined by \citet{2019MNRAS.485.2380M} for 158 stars that fall on the sample list of the two studies. The two luminosities  agree very well with a correlation coefficient, r, of 0.96.

\begin{figure}
 \centering
 \includegraphics[width=\columnwidth]{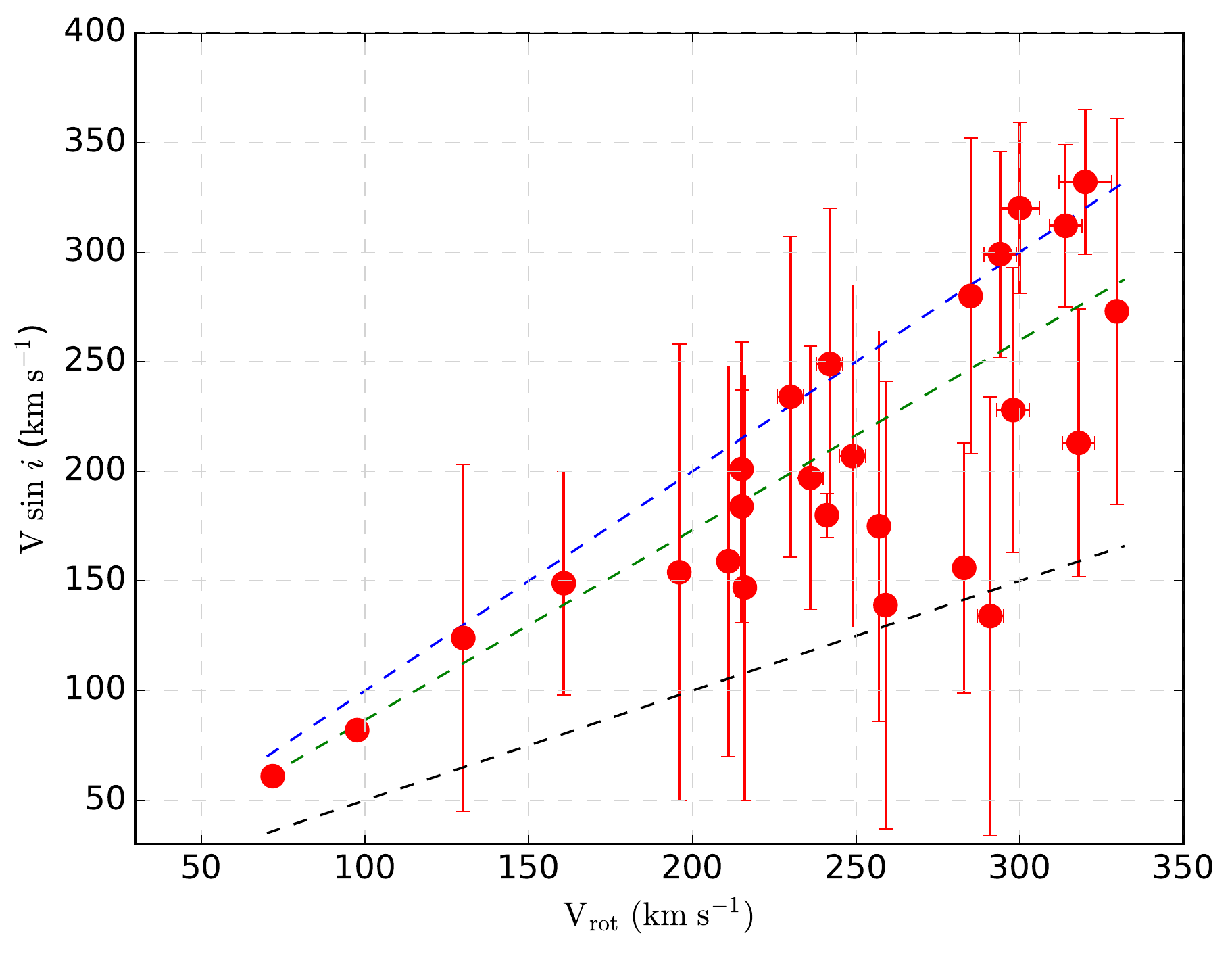}
 \caption{ The projected rotational velocity  as a function of equatorial rotational velocity. The blue, green and black dashed lines correspond to $i=90^\circ$, $60^\circ$ and $30^\circ$, respectively.} 
 \label{fig4}
\end{figure}

We calculated the average rotational velocity of Am/Fm and normal A stars as $\rm 105\pm 3~km~s^{-1}$ and $\rm 161\pm 3~km~s^{-1}$, respectively. Am/Fm stars rotate slowly relative to normal A stars. This confirms the conclusion from several studies \citep[e.g,][]{1995ApJS...99..135A,2004IAUS..224....1A,stateva, gebran16}. For 103 of the ``hump and spike'' stars in our sample, their projected rotational velocity ($\nu $~sin~$i$) is available from the LAMOST-{\it Kepler} project \citep{2016A&A...594A..39F} obtained from analysis of low resolution ($\rm \Re\simeq 1800$) spectra using the code ROTFIT \citep{2003A&A...405..149F, 2006A&A...454..301F}. For KIC\,9117875, KIC\,11443271 and KIC\,9349245, $\nu $~sin~$i$ determined from high-resolution spectroscopy by \citet{10.1093/mnras/stv528}, \citet{2017MNRAS.466.3060P} and \citet{2017MNRAS.470.2870N}, respectively, was adopted. Given the low resolution of LAMOST spectra, a value of $\nu $~sin~$i$ below $\rm 120~km~s^{-1}$ can not be determined. Therefore, for the stars tagged with ``$<$\,120'' in column\,11 of Tables~\ref{table1} and \ref{table_Am}, we only know that they are rotating slowly in the sense that their  $\nu $~sin~$i$ is below $\rm 120~km~s^{-1}$  \citep{2016A&A...594A..39F}. To validate the $\rm V_{rot}$ determinations, we compared our measurements with the available $\nu $~sin~$i$ from \citet{10.1093/mnras/stv528}, \citet{2016A&A...594A..39F}, \citet{2017MNRAS.466.3060P} and \citet{2017MNRAS.470.2870N}, but ignoring the stars with  $\nu $~sin~$i$\,$\rm< 120~km~s^{-1}$ by \citet{2016A&A...594A..39F}. The results are shown in Fig.~\ref{fig4}, where the blue, green and black dashed lines correspond to $i=90^\circ$, $60^\circ$ and $30^\circ$, respectively, and show the direction a star would move if it had a different inclination. Two stars appear slightly above the line sin\,$i=1.0$ (i.e; $i=90^\circ$). However, based on the lower limit of their error bars they fall in the reality zone. This implies that their $\nu $~sin~$i$ is slightly lower than what is observed. Generally, this is a promising result, though one would get assurance if $i$ was known and  $\nu $~sin~$i$ from high resolution spectroscopy was available for all the stars in our sample. Unfortunately, the majority of the stars in the sample are too faint ($\rm V\geq11~mag$) to be able to obtain a high-resolution spectrum with a sufficient quality with small (1-2\,m) ground-based telescopes. 

\begin{figure}
 \centering
 \includegraphics[width=\columnwidth]{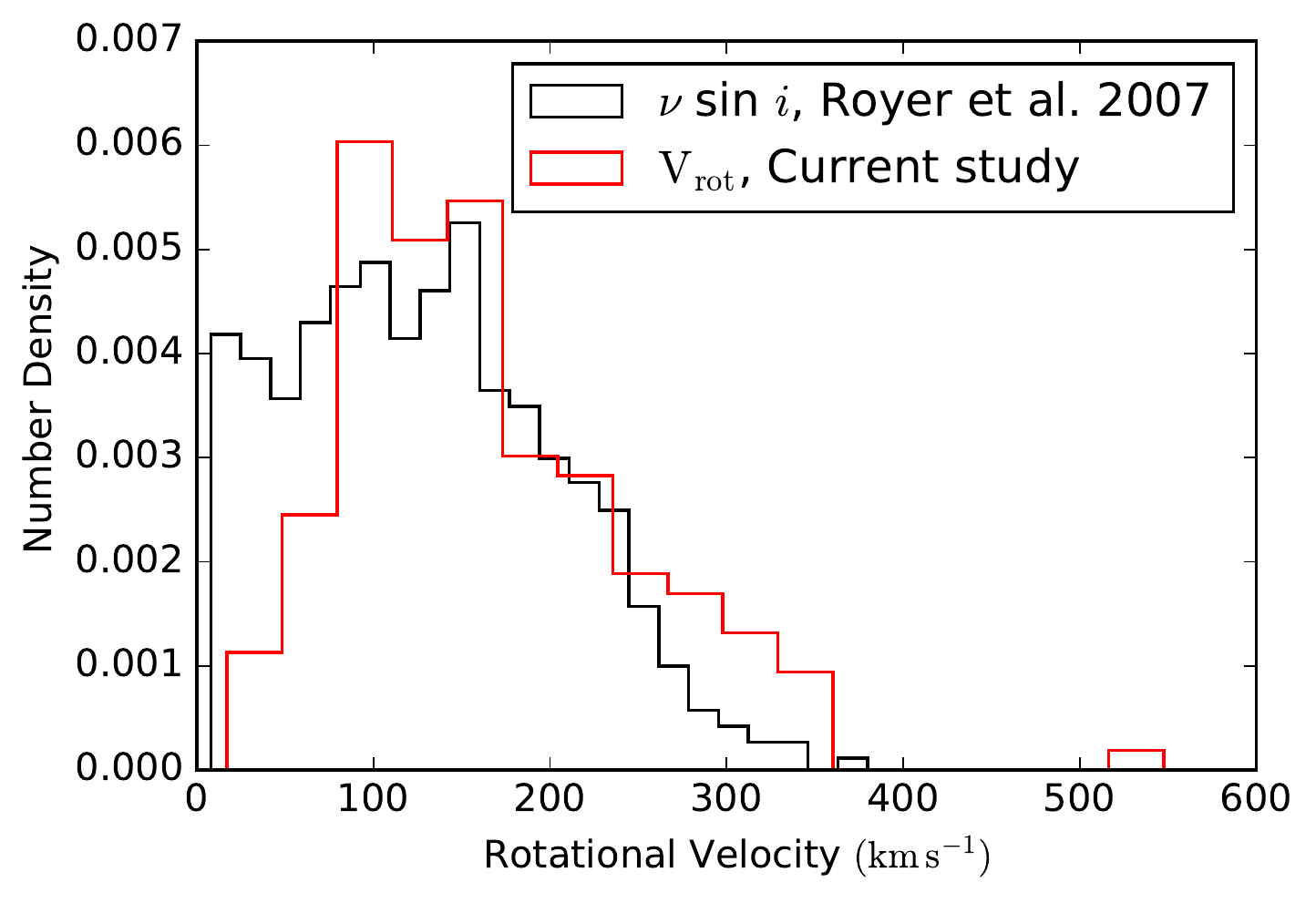}
 \caption{ The distribution the equatorial rotational velocity for the normal A stars in the current study and projected rotational velocity, $\nu $~sin~$i$ for A-type stars analysed by \citet{2007A&A...463..671R}.} 
 \label{fig:Royer}
\end{figure}

Fig.~\ref{fig:Royer} shows the distribution of the derived $\rm V_{rot}$ in the current study as compared with the $\nu $~sin~$i$ for the A-type stars studied by \citet{2007A&A...463..671R}. In their analysis, the authors considered only normal stars of spectral types from B9 to F2. Therefore, for homogeneity, only normal A stars in our sample are included. From Fig.~\ref{fig:Royer}, the normal A stars in our sample have a wide range of rotational velocity which is representative of A stars in general when compared with the distribution of $\nu $~sin~$i$ obtained by \citet{2007A&A...463..671R}. 

\subsection{Starspot size}
\label{sss}
The starspot size can give a clue of how strong the magnetic field is on the surface of a star. The knowledge of global magnetic field strength would facilitate giving informative and sound explanations about the development of the chemical peculiarities in intermediate-mass stars since the magnetic field may  stabilise against convection which would otherwise hinder atomic diffusion. From the light-curves in Fig.~\ref{fig2} and periodograms in Fig.~\ref{fig3}, amplitude variations and frequency spikes were observed. The amplitude variations in the light-curves and the spikes in the frequency spectra could be due to the presence of one or more starspots \citep{balona13, balona15, 2018MNRAS.474.2774S}. The starspot size can be determined directly from Doppler imaging \citep{1995MNRAS.275..534C, Barnes2002AN323..333B}. 
 
However, currently there is no direct method for calculating the starspot size photometrically. It has been reported that, within a magnetic cycle, the solar photometric variability increases with the activity levels \citep{Krivova2003}. Assuming that the largest active regions dominate the modulation, as a proxy, the photometric amplitude of the rotation frequency can be extrapolated to be representative of the starspot size \citep{balona13, 2017MNRAS.472.1618G}. The amplitude may not give the exact magnitude of the starspot size but gives a good approximation.

From the frequency spectra, we obtained and considered the amplitudes of the spikes to be the photometric amplitudes (${\rm A_{rot}}$) associated with rotation. The uncertainties in amplitude were calculated using Monte Carlo simulations \citep{2005CoAst.146...53L}. The results are listed in Tables~\ref{table1} and \ref{table_Am}.  The average ${\rm A_{rot}}$ of Am/Fm and normal A stars was found to be $\rm \sim 21\pm2 ~ ppm$ and $\rm \sim 19\pm2~ppm$, respectively.

\begin{figure}
 \centering
 \includegraphics[width=\columnwidth]{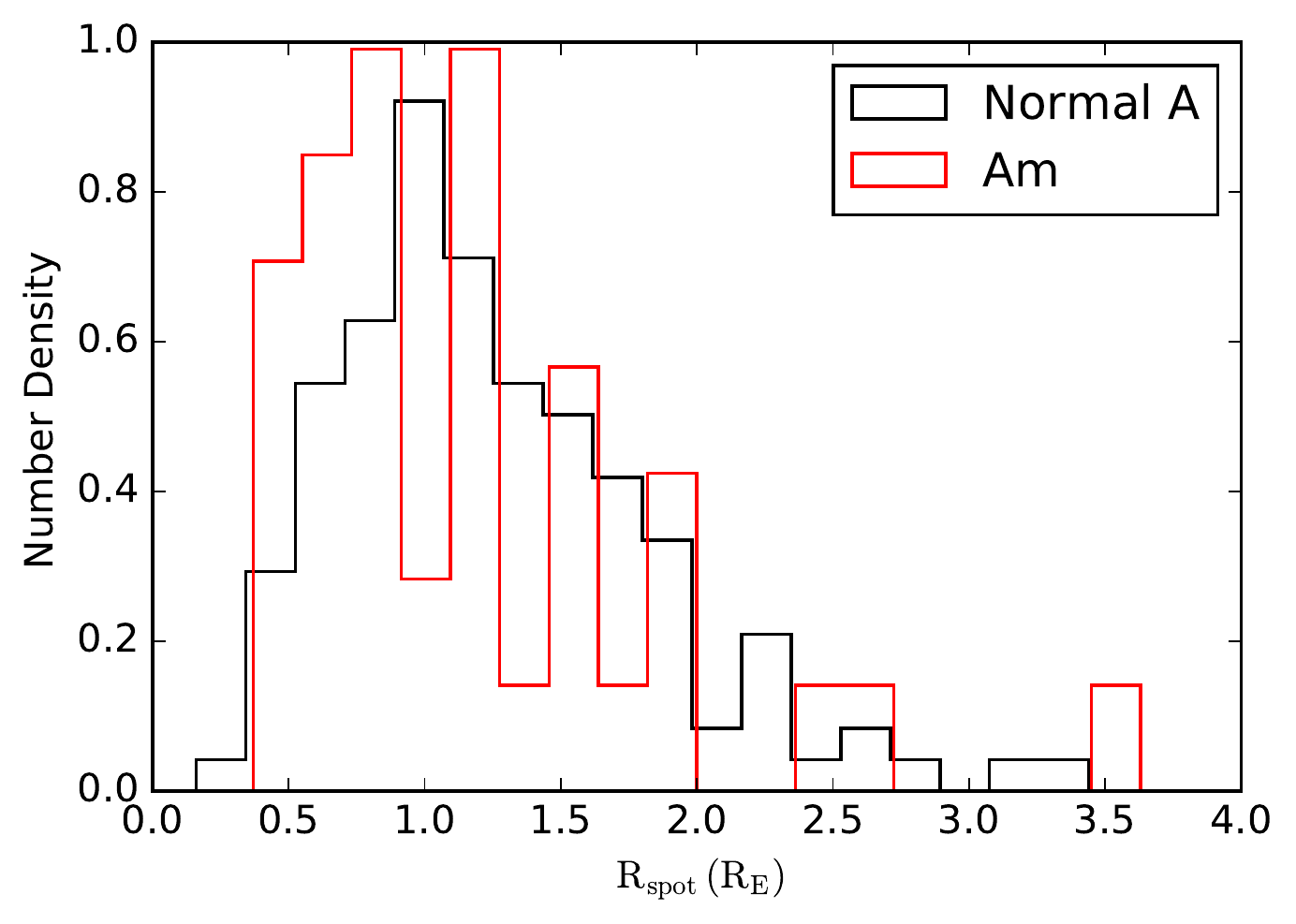}
 \caption{ The distribution of the spot size for normal A and Am/Fm stars in our study calculated using Eq.\,\ref{rspot2}.}
 \label{fig5}
\end{figure}

For simplicity, we assume a single, circular and black spot to produce the amplitude of the spike and the stars to be spherical. This is identical to determining the size of an exoplanet from a transit. The radius of the spot ($\rm R_{spot}$) is given by the standard relation,
\begin{equation}
 \rm 
 R_{spot}=\sqrt{\rm A_{rot}} R.
 \label{rspot1}
\end{equation} 
 We estimated the spot radii using Eq.\,\ref{rspot1}, the results are given in column\,12 of Tables~\ref{table1} and \ref{table_Am} and their distribution is shown in Fig.~\ref{fig5}. Indeed, the spots maybe regions of a lower $T_{\rm eff}$ \citep{Strassmeier2009} and hence could appear a little darker instead of completely dark. Considering the $T_{\rm eff}$ of the active region (spot) and the photosphere of the star to be $T_{\rm spot}$ and $T_{\rm phot}$, respectively, Eq.\,\ref{rspot1} becomes,
\begin{equation}
 {\rm
 R_{spot}=\sqrt{\rm A_{rot}} R}\left(\frac{T_{\rm phot}}{T_{\rm spot}}\right)^2.
 \label{rspot2}
\end{equation}
\cite{Strassmeier2009} reported that the spot-to-photosphere temperature ratio ($T_{\rm spot}/T_{\rm phot}$) is about 0.8 for active G and K stars and that, on average, the temperature difference appears to be larger for hotter stars. Therefore, we expect $T_{\rm phot}/T_{\rm spot}\geq 1.25$, which translates via Eq.\,\ref{rspot2} into starspot size that is larger by a factor of $\geq 1.56$ \citep{Strassmeier2009}.
There is a theoretical evidence that spots originating from ultra-weak magnetic fields might be hotter with a temperature difference of the order of 10\,K, and hence only a very small effect on the starspot size  \citep{2019ApJ...883..106C}. In addition, the calculated $\rm R_{spot}$ is the starspot group size if there are several spots on the star. 
The distributions of both normal A and Am/Fm stars are similar and range from 0.2 to about 3.7~$\rm R_E$, where $\rm R_E$ is the Earth radius. The derived average spot radius is $\rm 1.01\,\pm\,0.13$ and $\rm 1.16\,\pm\,0.12$\,$\rm R_E$ for Am/Fm and normal A stars, respectively. Generally, the starspot size in normal A and Am/Fm stars is smaller than previously estimated by \citet{balona13} and \citet{balona15} by about 38\,\%. This implies that the activity in Am/Fm and normal A stars is weaker than previously suggested by these studies. 

\subsection{Decay-time scale}
The starspot decay-time scale gives a clue of how long the overall starspots are sustained on the stellar surface, once they have appeared. The rate of starspot growth and decay could be linked to the surface magnetic field strength. The starspot growth and decay contributes to amplitude variations in the light-curves. For a better understanding and explanation of various phenomena observed in normal A and Am/Fm stars, good estimates of the decay-time scale are needed. The decay-time scale was determined from fitting autocorrelation functions (ACFs) of the light-curves. The ACFs show how similar light-curves are to themselves at certain time differences.

\begin{figure}
  \centering
 \includegraphics[width=\columnwidth]{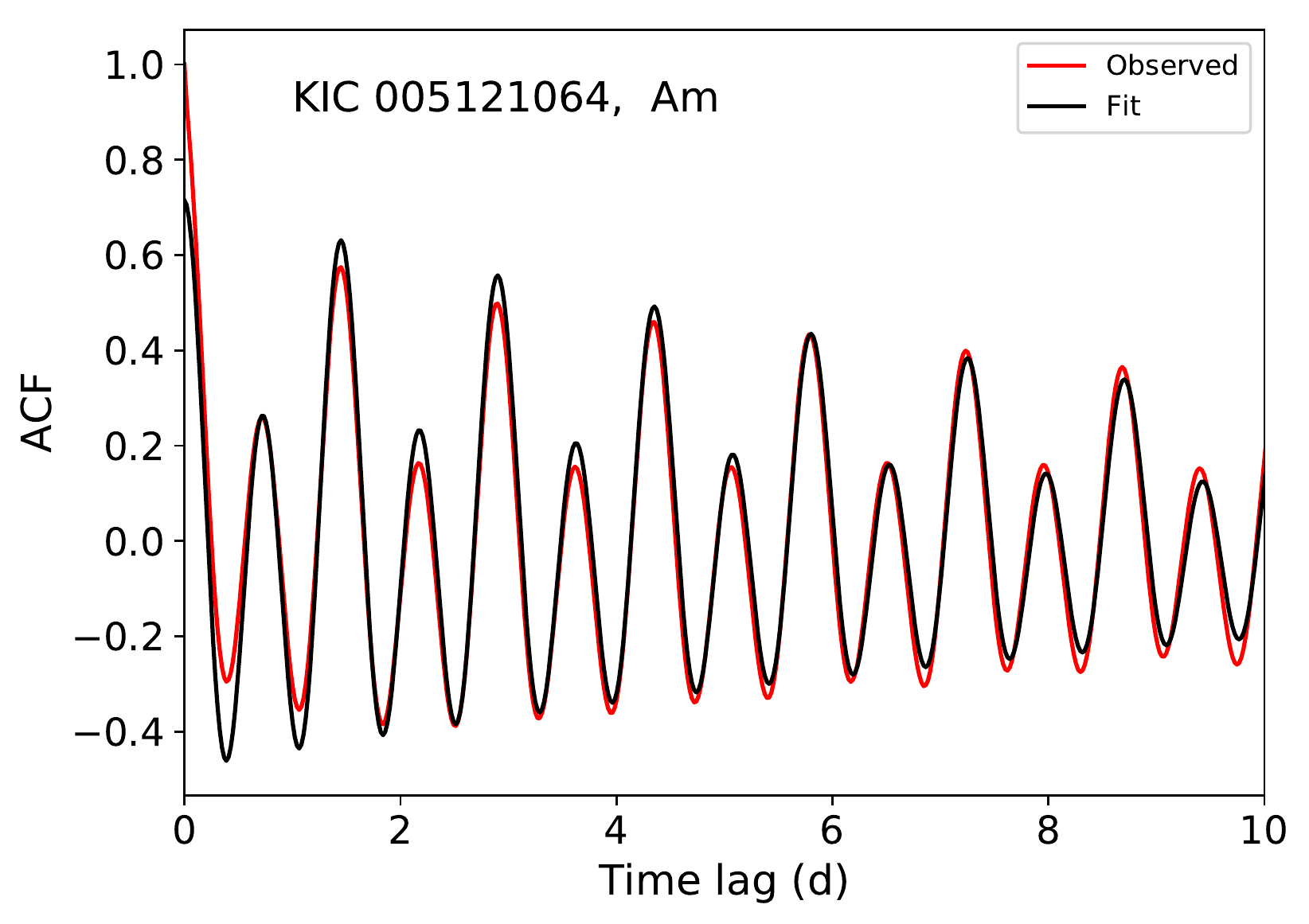}
 \caption{A typical example of autocorrelation function where the second term of Eq.~\ref{uSHO} is manifested.}
 \label{fig6}
\end{figure}

\begin{figure}
  \centering
 \includegraphics[width=\columnwidth]{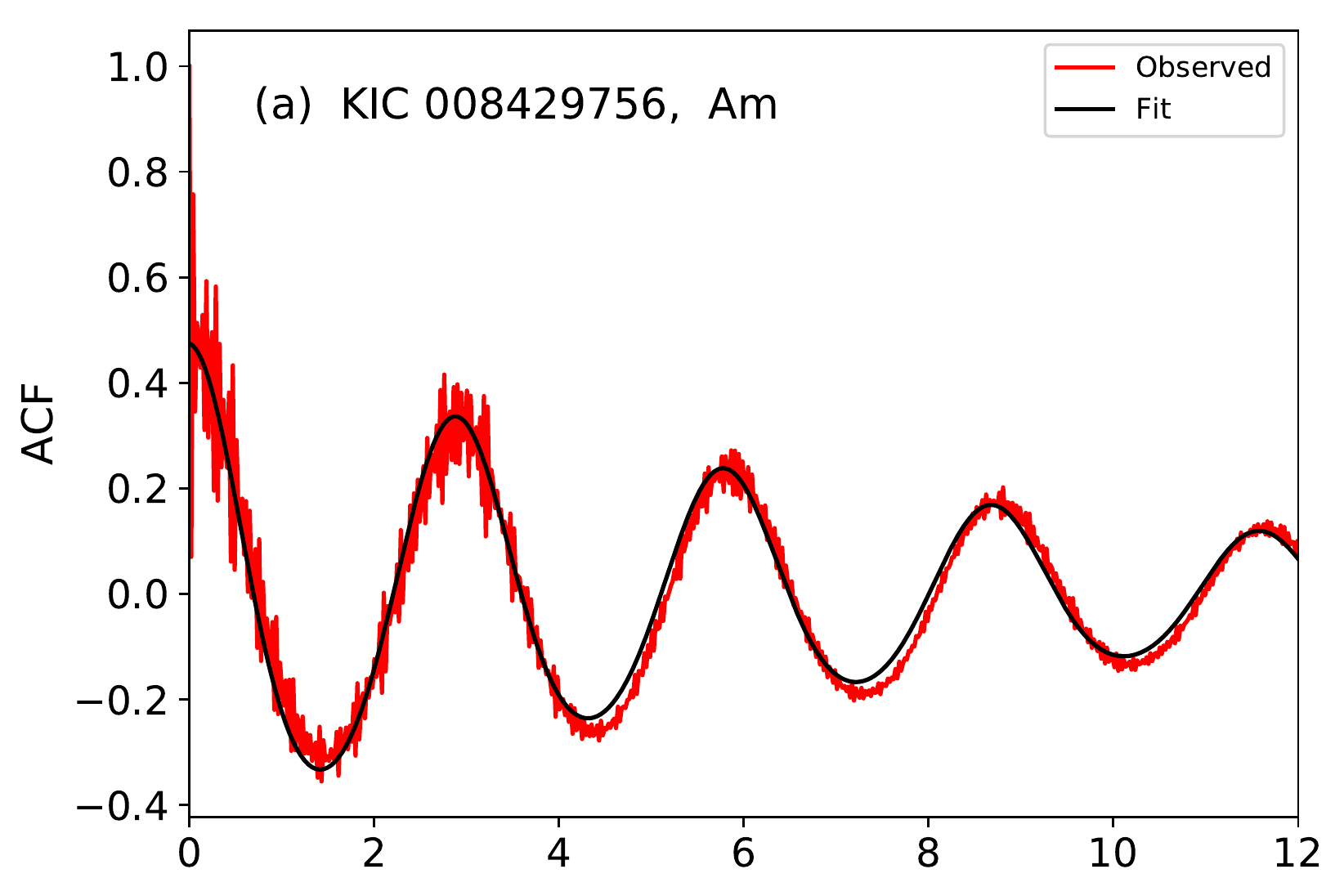}
  \includegraphics[width=\columnwidth]{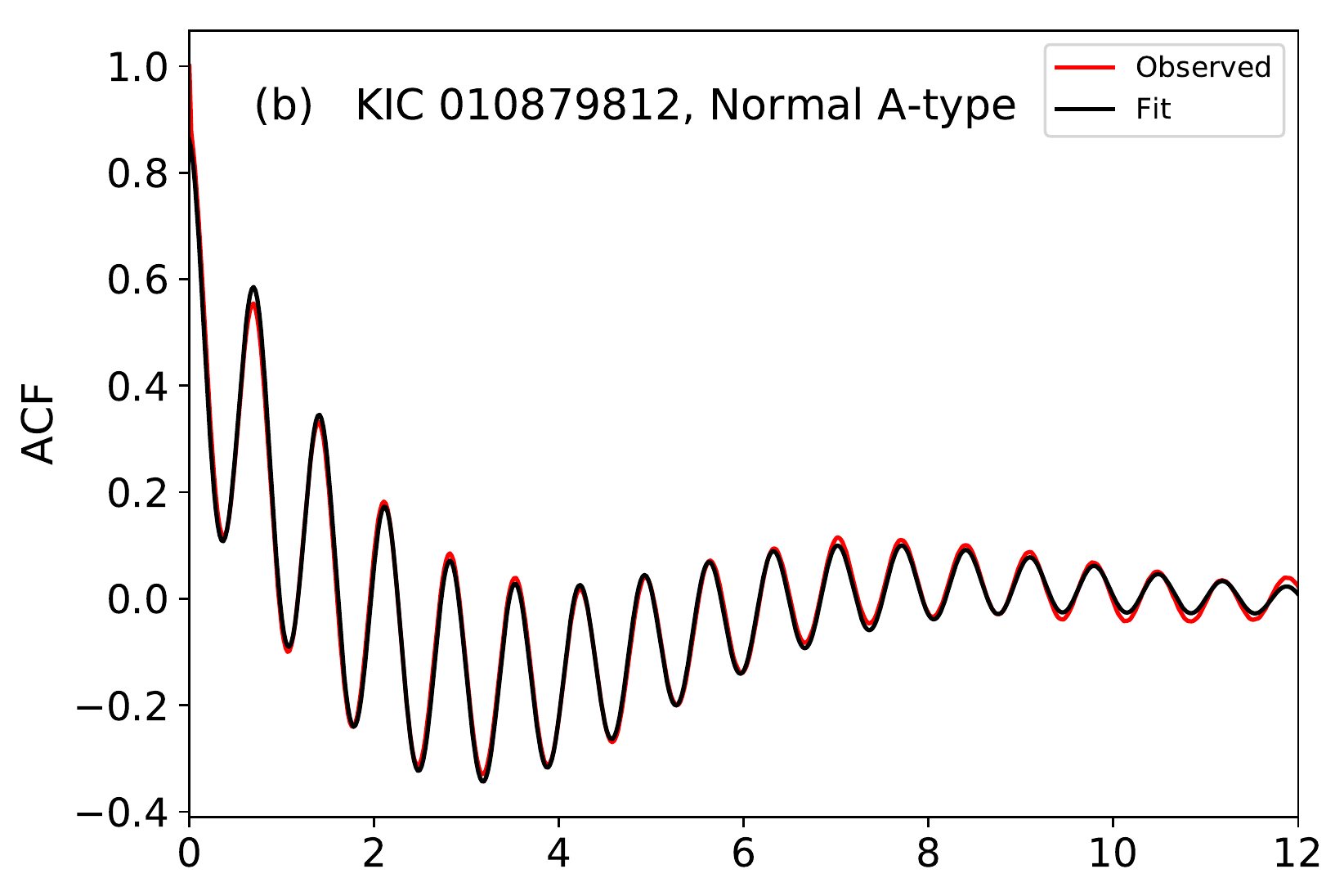}
 \includegraphics[width=\columnwidth]{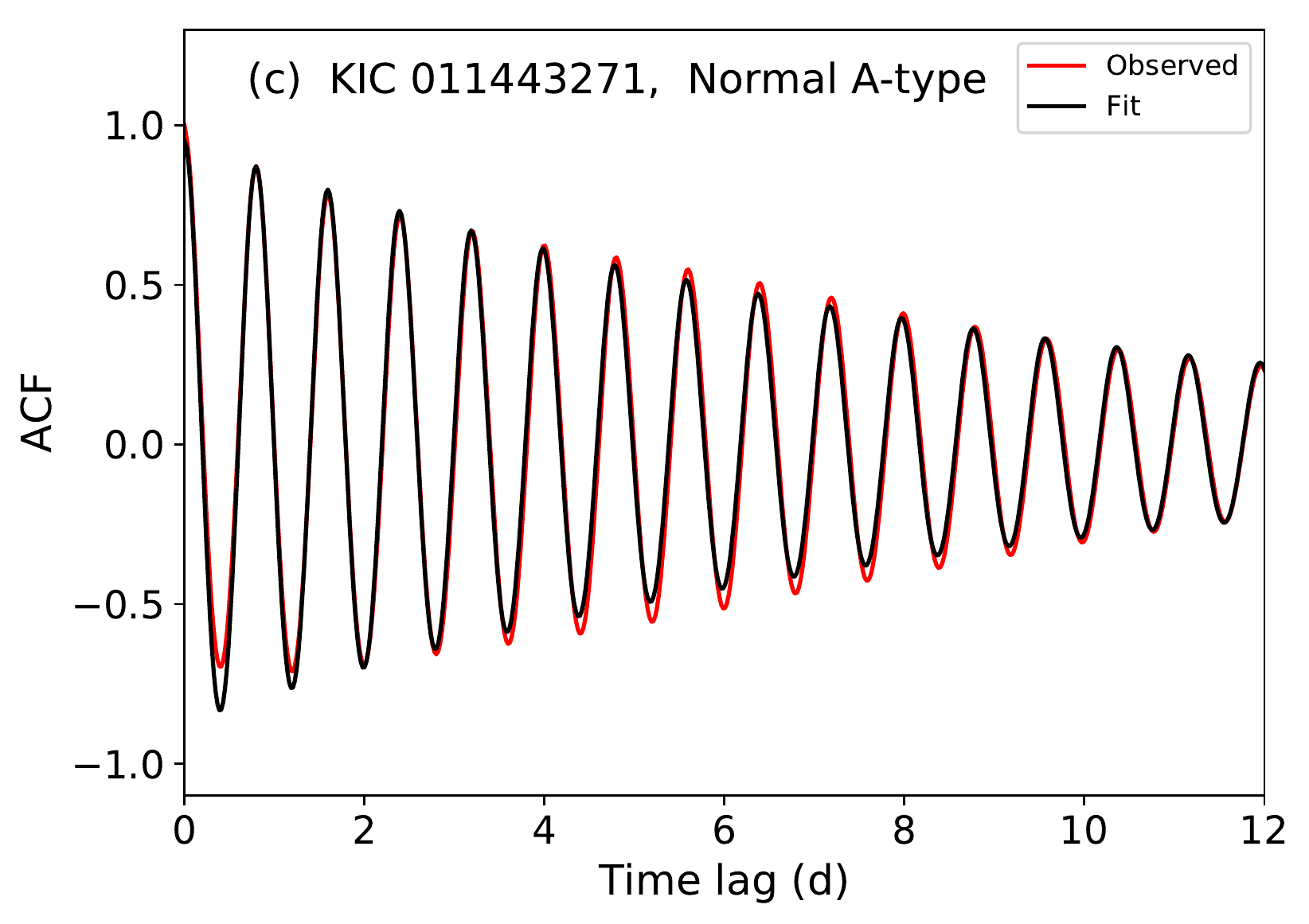}
 \caption{Examples of autocorrelation functions (ACF) plots where the spot(s) is(are) dominant on only one side of the star and so the second term of Eq.~\ref{uSHO} is overshadowed. The first peak corresponds to a time lag equivalent to the rotation period.}
 \label{fig7}
\end{figure}

For each star, the {\it Kepler} light-curves from all the available quarters were combined together end-to-end and the resultant  light-curve was cross-correlated with itself at a given time lag to produce ACF \citep{2013MNRAS.432.1203M, 2014ApJS..211...24M}. The ACF was normalised on dividing it by the sum of deviations in the flux. For the majority of the stars, the ACFs increase and decrease quasi-sinusoidally depending on the presence of a co-rotating obstacle in the field of view of the star.  At time lags greater than zero, the ACF tends to mimic the displacement of an under-damped simple harmonic oscillator (uSHO), \citep{2017MNRAS.472.1618G}
\begin{equation}
 \rm y(t)=e^{-t/\tau_{DT}}\left( Acos\left(\frac{2\pi t}{P_{ACF}}\right)+Bcos\left(\frac{4\pi t}{P_{ACF}}\right)+y_0\right),
 \label{uSHO}
\end{equation} where
\begin{equation}
 \rm t= \Delta T \times n,
\end{equation} here t is the time lag in days, $\rm \Delta T$ is the median time difference of the light-curve, n ascends from 0 to the total number of ACFs, y(t) is the ACF, $\rm \tau_{DT}$ is the decay-time scale of the dominant active region, $\rm P_{ACF}$ is the stellar rotation period, and A, B and $\rm y_0$  do not represent any physical stellar properties but are constants that are useful in fitting the uSHO equation. All the ACFs were fitted with Eq.~\ref{uSHO} using $\rm\chi^2$ minimisation. An example of the ACF for the Am/Fm star KIC\,5121064 is given in Fig.\,\ref{fig6}. The second term in Eq.~\ref{uSHO} is significant for this star, inducing subsequent high and low local maxima in the ACF. Indeed, in Fig.~\ref{fig6}, the second local maximum corresponds to the rotation period ($\rm P_{ACF}$) while a lower local maximum is observed at almost half the rotation period. This smaller peak is catered for by the second term in Eq.~\ref{uSHO}. An ACF as the one in Fig.~\ref{fig6} indicates that there is(are) weak starspot(s) opposite to the stellar face with the dominant starspot(s).
Examples of autocorrelation functions where this scenario is not pronounced are shown in Fig.~\ref{fig7}. Inspection of the ACFs was done visually. In general, good quality fits were produced by the least-squares routine ($\rm \chi^2$ minimisation).

\begin{figure}
  \centering
  \includegraphics[width=\columnwidth]{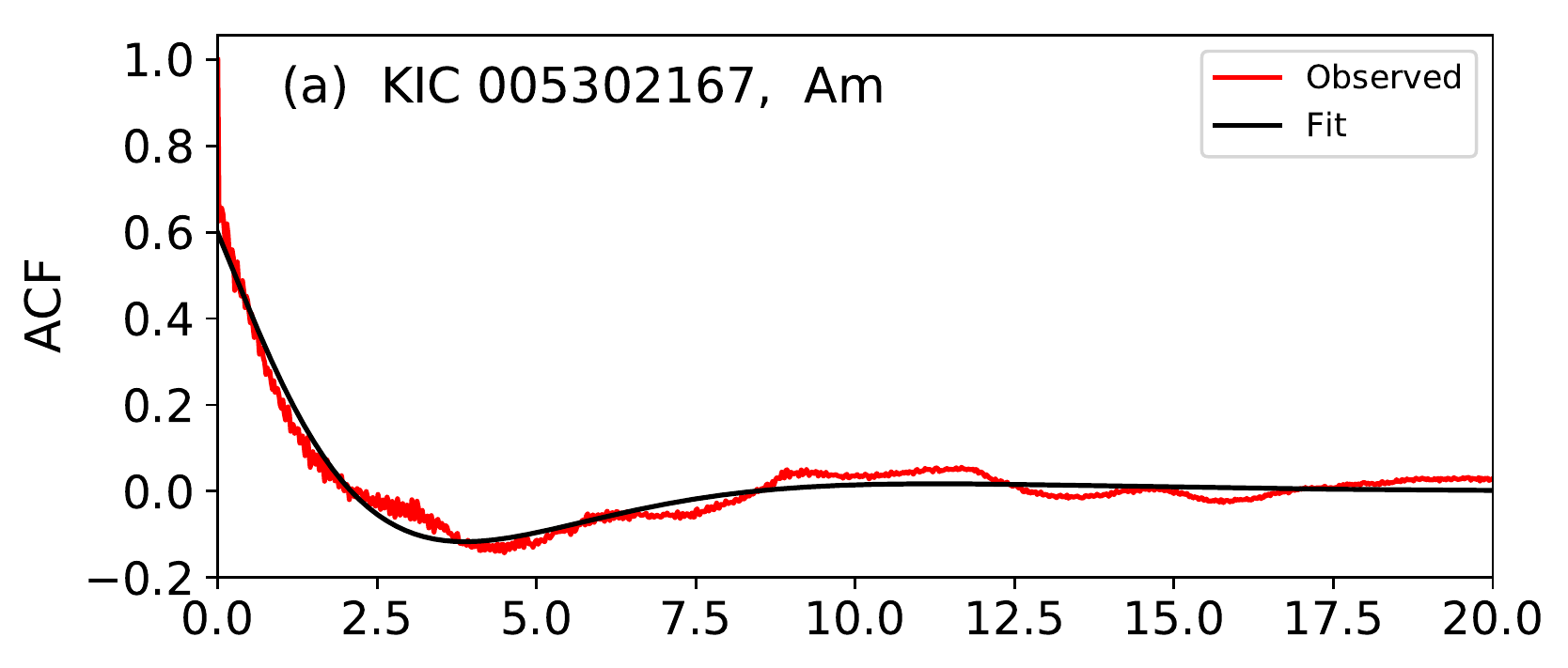}
  \includegraphics[width=0.95\columnwidth]{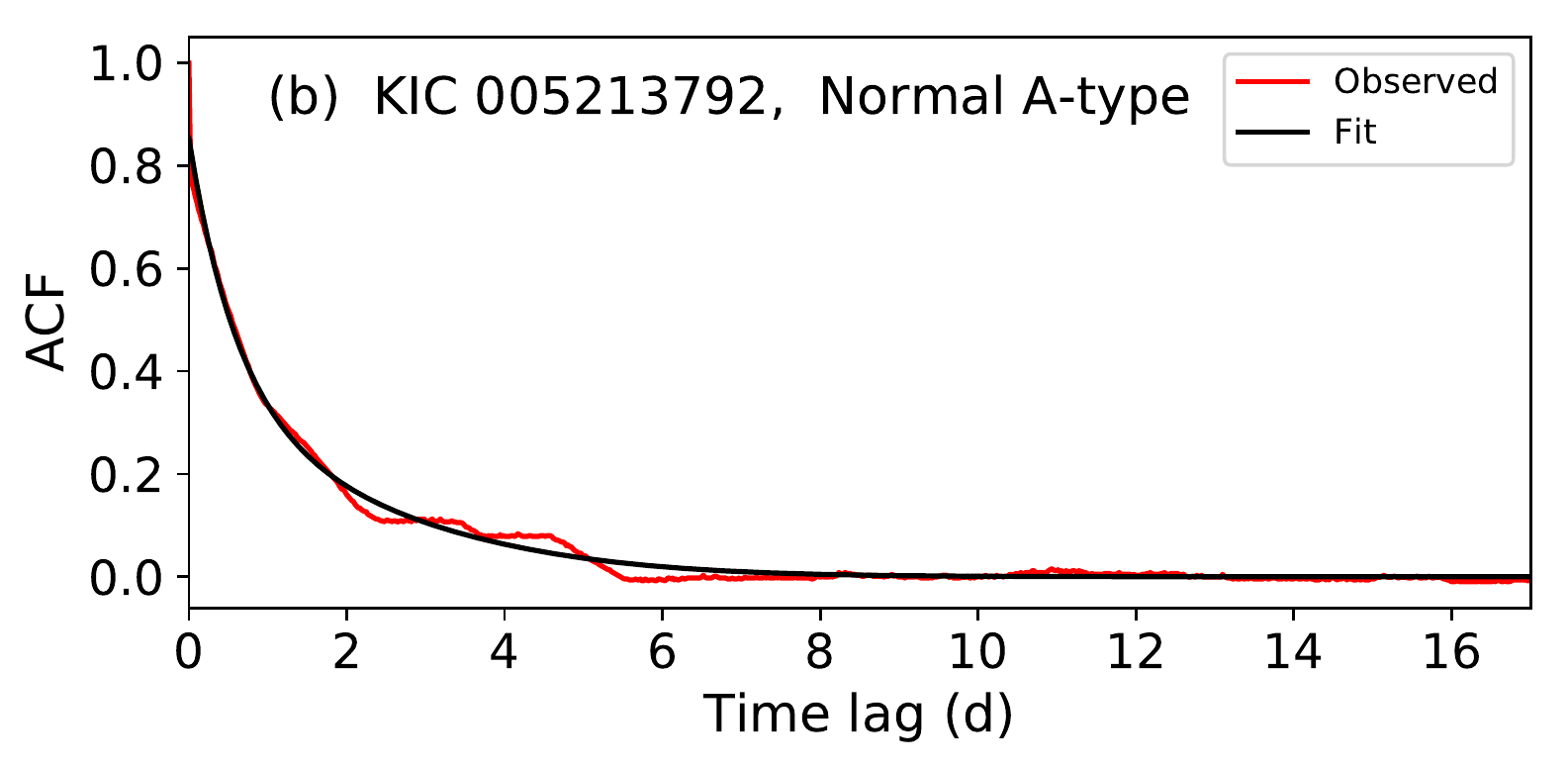}
 \caption{Examples of autocorrelation functions (ACF) which do not show any significant peaks.}
 \label{fig8}
\end{figure}

\begin{figure}
  \centering
 \includegraphics[width=\columnwidth]{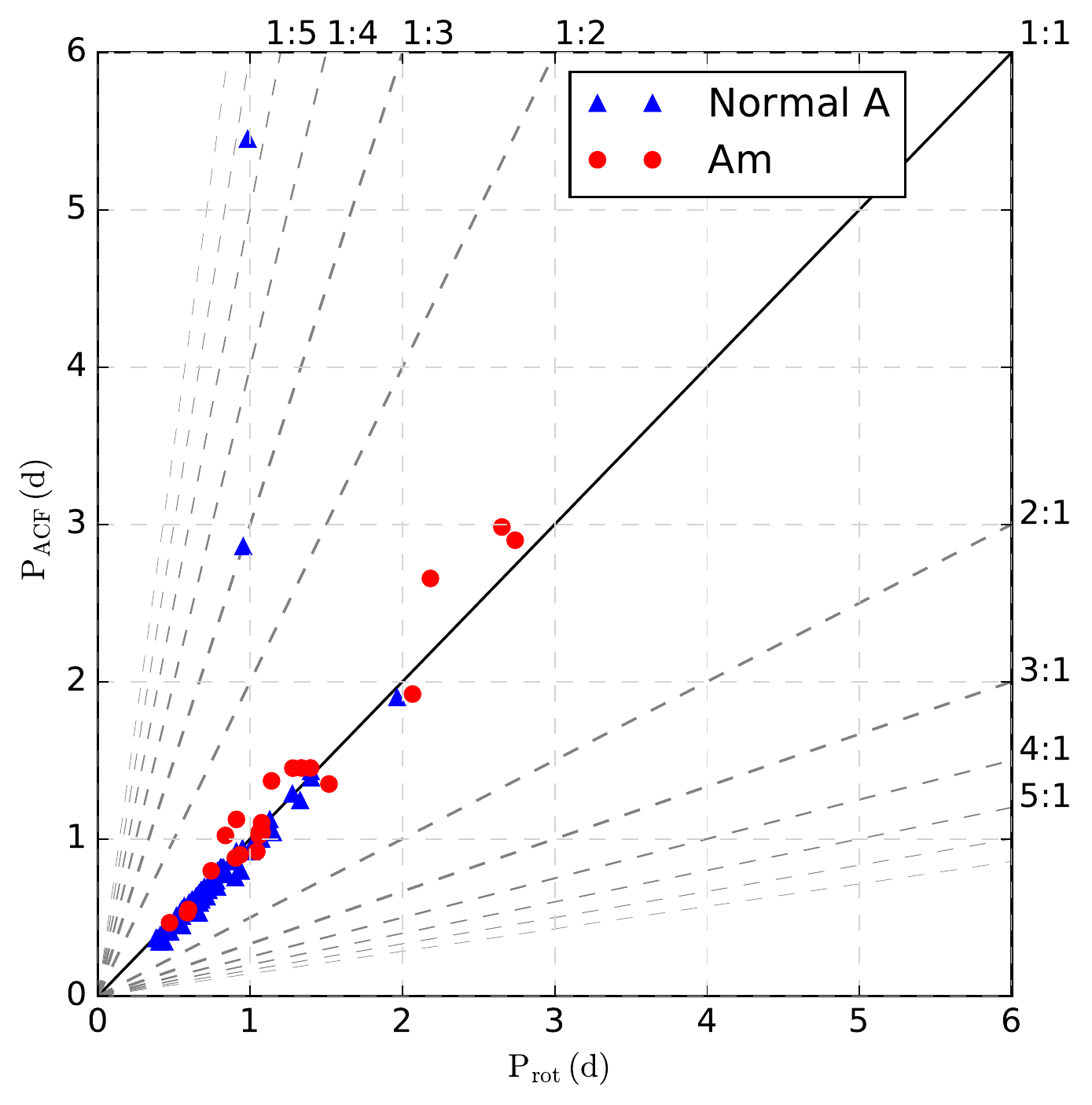}
 \caption{Comparison of rotation periods derived by different methods. The black solid line represents 1:1 relation, while the dashed lines indicate the harmonics (1:N) and subharmonics (N:1). There are two normal A stars with periods that differ a factor of 3 or more.}
 \label{fig6b}
\end{figure}

About 40\% of the ACFs do not show any significant peaks. Examples of such ACFs are shown in Fig.~\ref{fig8}, where it is observed that generally the ACF decays instantly without significant local maxima. This points towards a short decay-time and low amplitudes, so it suggests very weak or the absence of starspots and co-rotating structures. However,  the ``hump and spike'' features exist in their frequency spectra.  This implies that significantly large starspots may not be required to produce the existing spikes in the frequency spectra. A similar observation was made by \citet{2018MNRAS.474.2774S} as very small starspots were enough to produce spikes in the models. The resulting values of the decay-time $\rm \tau_{DT}$ and the  period $\rm P_{ACF}$  are listed in Tables~\ref{table1} and \ref{table_Am}. The stars without any significant peaks in their ACFs have (*) assigned to their $\rm \tau_{DT}$ values in the Tables and they were not considered during  the discussion of the decay-time scale.  In Fig.~\ref{fig6b}, rotation periods derived from ACFs ($\rm P_{ACF}$) are compared with periods from the spikes ($\rm P_{rot}$). The black solid and dashed lines represent the 1:1 relation and the harmonics (1:N) and subharmonics (N:1), respectively. The two periods are well in agreement. However, there are two normal A stars (KIC\,4557097 and KIC\,10068389) for which the periods of both methods do not agree. This implies that the spikes get overshadowed by the dominant frequencies during the ACF analysis. This is because the spikes are not the dominant peaks in their periodograms. This also applies to all the stars that do not perfectly lie along the line 1:1. We also never considered the two stars (KIC\,4557097 and KIC\,10068389) in the analysis of the decay-time scale of the sample.

The majority of the stars have decay-time scales of a few days or less, as shown in Fig.~\ref{fig9}(e$_{1-3}$). The average decay-time in Am and normal A stars is $3.6\pm0.2$ and $1.5\pm0.2$ days, respectively. The difference may not be associated with the difference in the rotation velocities between the two groups of stars (discussed in Sect.\,\ref{vel}), as we observed no correlation between the two parameters. However, the difference could be attributed to the difference in effective temperature. The Am/Fm stars are, in general, cooler than normal A stars in our sample as shown in Fig.~\ref{Teff:dist}. The thin convective sub-surface layer reduces as the effective temperature increases, which affects the decay lifetime of the available spots. The decay-time scales in Am and normal A indicate that spots in these stars are short-lived compared with those in G, K and M stars. Spots of similar emergence and disappearence time of only a few days were observed in Vega by \cite{2017MNRAS.472L..30P}. The spots in G, K and M stars can stay on the surface for as long as 400 days \citep{2017MNRAS.472.1618G}. \citet{2017MNRAS.472.1618G} observed that big starspots live longer on any given star. In Fig.~\ref{figDT}, there is a weak positive correlation between spot radius ($\rm R_{spot}$) and the decay-time scale ($\rm \tau_{DT}$) for the stars in our sample. This implies that the $\rm \tau_{DT}$-values weakly depend on the spot size in these stars, and a significantly large sample is required to confirm the observation. \citet{2017MNRAS.472.1618G} also reports that starspots decay more slowly on cooler stars. Normal A and Am/Fm stars are hotter than G, K and M stars, which verifies the observations of short decay-time scale. The observed spot sizes and decay-time scales suggest the presence of a weak magnetic field in the stars of our sample. The magnetic fields could be in the similar range as observed for Vega \citep{2009A&A...500L..41L}, Sirius \citep{2011A&A...532L..13P}, $\beta$ UMa and $\theta$ Leo \citep{blaz2}, HD~188774 \citep{2015MNRAS.454L..86N}, Alhena \citep{blaz} and $\rm \rho~Pup$ \citep{ 2017MNRAS.468L..46N}. \cite{2019ApJ...883..106C} reported a similar suggestion that, except for the Ap stars, the other main sequence A-type stars can have magnetic fields of a few Gauss. Some studies show that such magnetic field can support weak starspots \citep[e.g;][]{blaz, 2017MNRAS.468L..46N,2019ApJ...883..106C}. Nevertheless, we may not rule out the possibility of the observed spots having an origin other than the magnetic fields that is still unknown.

\subsection{Correlation of rotation and spot properties with other parameters}
\label{discuss}
\begin{figure*}
  \centering
  \includegraphics[width=0.8\textwidth]{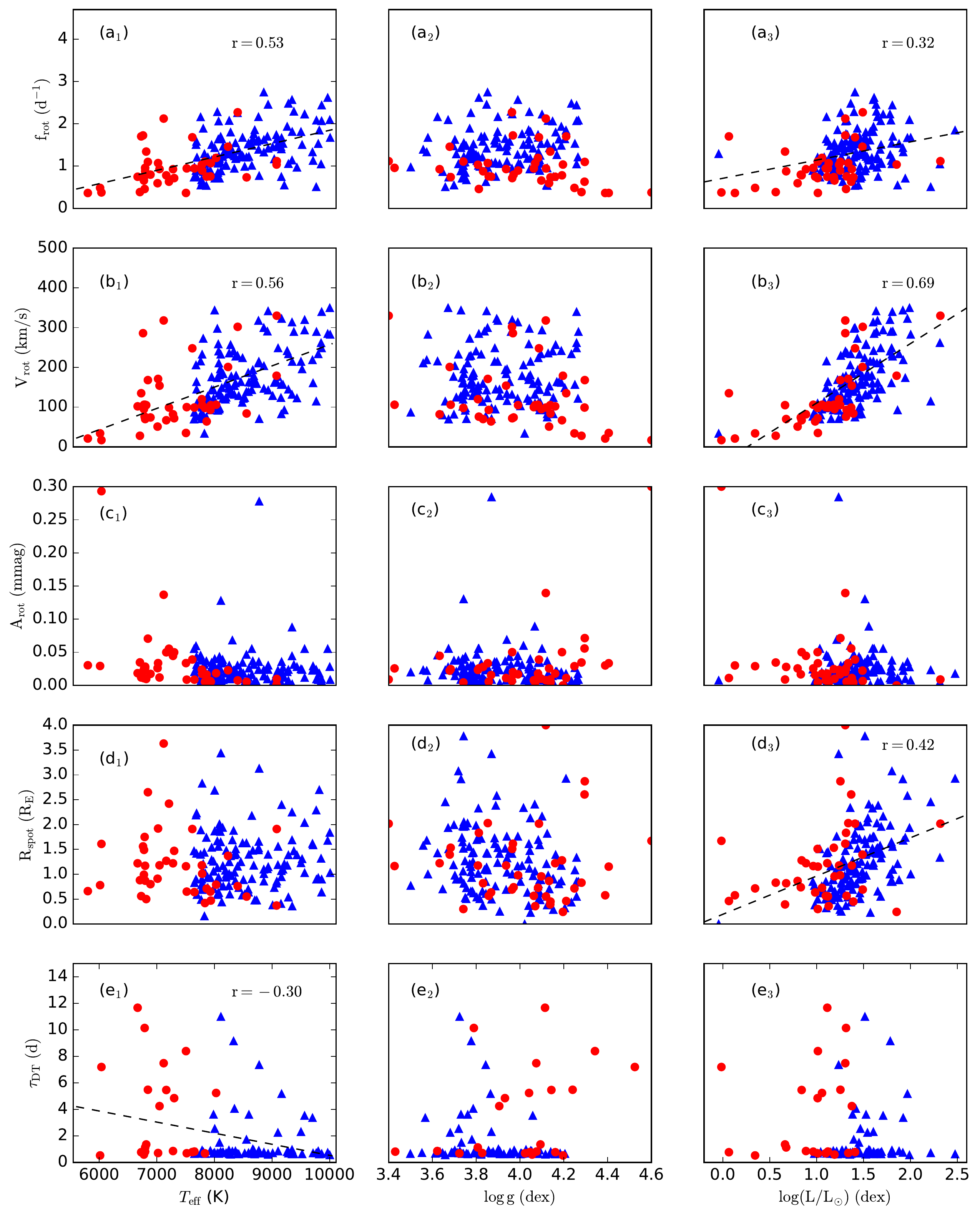}
 \caption{The rotation frequency (a$_{1-3}$), rotational velocity (b$_{1-3}$), photometric amplitude (c$_{1-3}$), spot radius (d$_{1-3}$) and starspot decay life-time (e$_{1-3}$) versus effective temperature, surface gravity and luminosity. The $\rm T_{eff}$ and log\,g-values from the revised catalog of {\it Kepler} targets for the $\rm Q_{1-17}$ (DR25) \citep{2017ApJS..229...30M} were adopted. The red dots represent Am/Fm stars and blue triangles are normal A stars. The black dashed lines indicate the fits between the respective parameters and the value of the correlation coefficient (r) is given in the top right corner.}
 \label{fig9}
\end{figure*}

\begin{figure}
 \centering
 \includegraphics[width=\columnwidth]{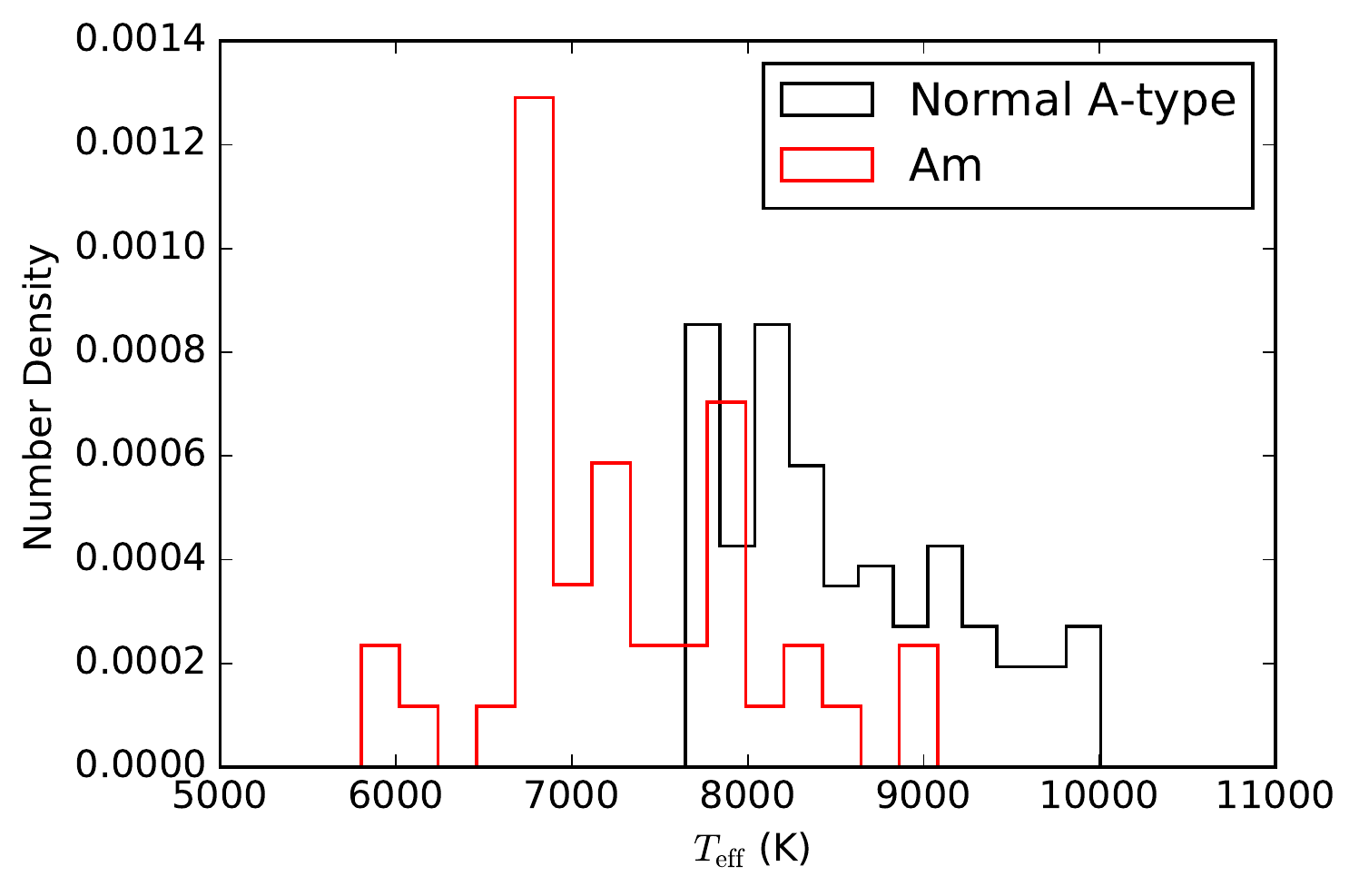}
 \caption{ The distribution of the effective temperature for normal A and Am/Fm stars in our study.}
 \label{Teff:dist}
\end{figure}

\begin{figure}
  \centering
 \includegraphics[width=\columnwidth]{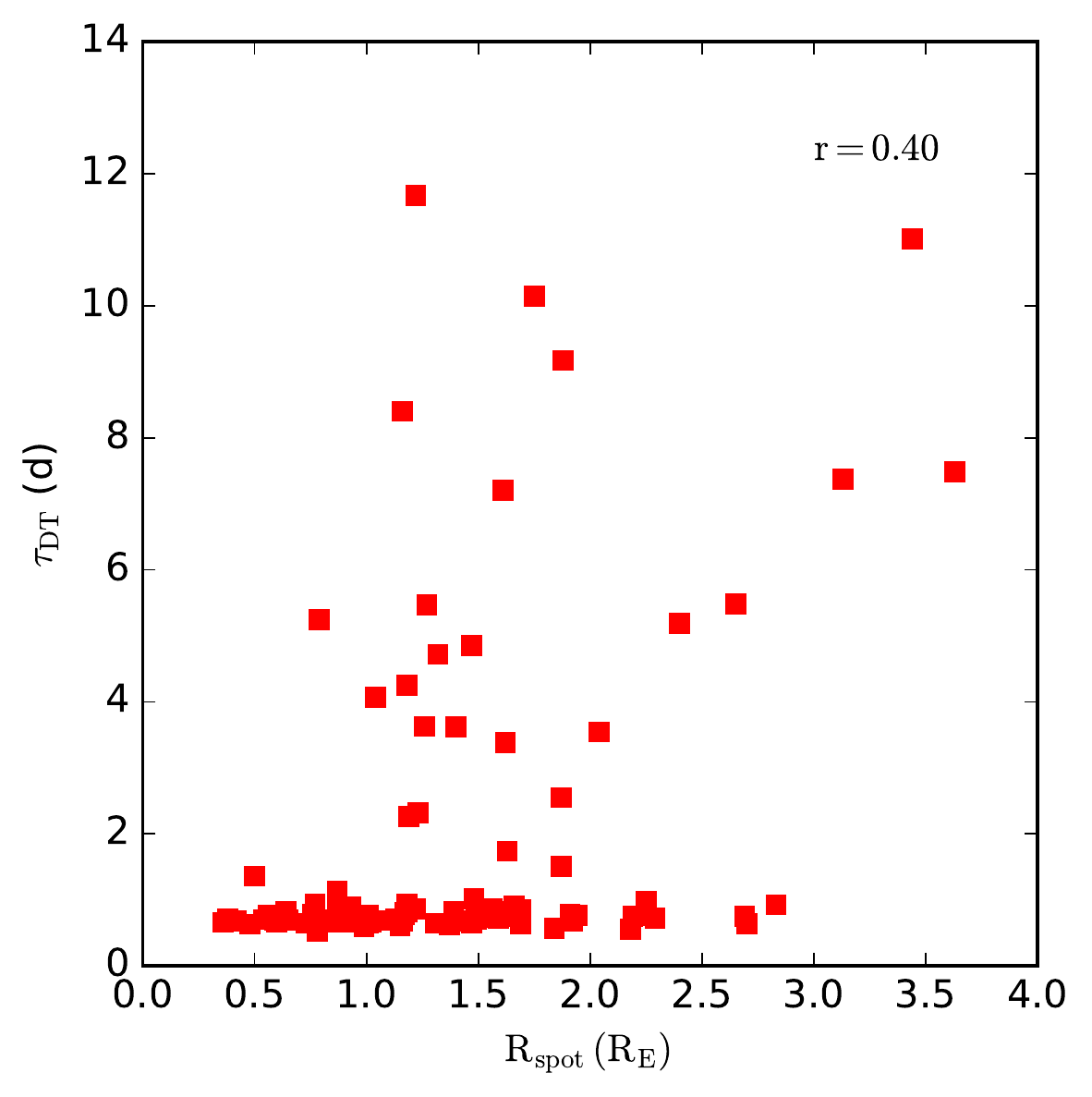}
 \caption{The correlation between the spot radius ($\rm R_{spot}$) and the decay-time scale ($\rm \tau_{DT}$) for normal A and Am/Fm stars. The black dashed line shows the fit. r-value the correlation coefficient is given at the top right corner.}
 \label{figDT}
\end{figure}

This study presents a detailed analysis of the rotation properties of ``hump and spike'' stars. We analysed {\it Kepler} photometric data of 170 stars. The distribution of the rotation properties of our sources over the temperature ($\rm T_{eff}$), surface gravity (log\,g) and luminosity (log(L/L$_\odot$)) domains is represented in Fig.~\ref{fig9}. Panels (a$_{1-3}$), (b$_{1-3}$), (c$_{1-3}$), (d$_{1-3}$) and (e$_{1-3}$) of Fig.~\ref{fig9} show how $\rm f_{rot}$, $\rm V_{rot}$, $\rm A_{rot}$, $\rm R_{spot}$ and $\rm \tau_{DT}$, respectively, vary with the $\rm T_{eff}$, log\,g and  log(L/L$_\odot$) of the studied stars. From panels (a$_1$) and (b$_1$), rotation frequency and velocity are moderately correlated with effective temperature. The $\rm A_{rot}$ (panel (c$_1$)) and spot radius (panel (d$_1$)) have no significant relation with $\rm T_{eff}$ for these stars. A weak dependence of $\rm \tau_{DT}$ on $\rm T_{eff}$ is observed in panel (e$_{1}$). A similar observation is reported by \citet{2017MNRAS.472.1618G} for the G, K and M stars. From the panels in the middle, the rotation and spot properties are independent of log\,g for these stars. Based on the r-values in panels (a$_{3}$) and (d$_{3}$), the rotation frequency and the spot size weakly depend on luminosity. Panel (b$_{3}$) indicates a moderately strong relation between $\rm V_{rot}$ and log(L/L$_\odot$). There is no dependence of $\rm A_{rot}$ and $\rm \tau_{DT}$ on log(L/L$_\odot$) for these stars. From panels (c$_{1-3}$), (d$_{1-3}$) and (e$_{1-3}$), we conclude that in general, the photometric amplitudes, spot radii and spot decay-time scales of normal A and Am/Fm stars are on average 0.025\,mmag, 1.61\,$\rm R_E$ and 3.3\,days, respectively.
  
\begin{figure}
  \centering
  \includegraphics[width=\columnwidth]{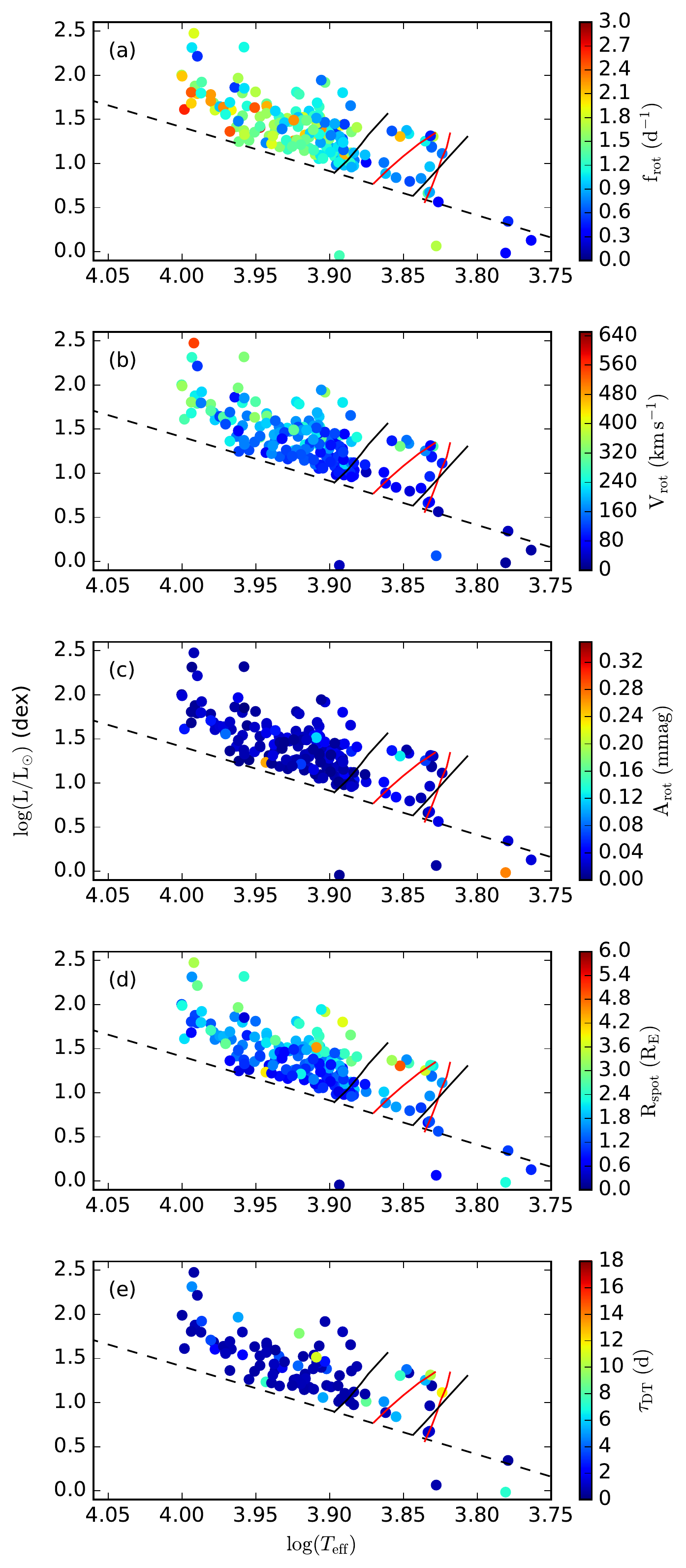}
 \caption{The distribution of the rotation and spot properties in the Hertzsprung-Russell diagram. The black dashed line represents the zero age main sequence. The black and red solid lines indicate the  theoretical $\rm \delta$\,Scuti and $\gamma$\,Doradus instability strips, respectively, as calculated by \citet{2004A&A...414L..17D}.}
 \label{HRD}
\end{figure}

Fig.~\ref{HRD} shows distribution of the rotation and spot properties in the Hertzsprung-Russell diagram. The majority of the stars are beyond the blue edge of the theoretical $\delta$\,Scuti instability strip calculated by \citet{2004A&A...414L..17D}. In Fig.~\ref{HRD}(a) and (b), the rotation frequency and velocity increase with towards the more massive stars, while the spot radius, in panel (d), seem to increase as stars move off the zero age main sequence. There are no observable patterns for photometric amplitude and decay-time scale in  panels (c) and (e), respectively.  The observations will be confirmed when a significantly large number of stars is used. Efforts are being made to search for ``hump and spike'' stars in the K2 and {\it TESS} archives.
 
The magnetic field in all the stars in the sample is presumed ultra weak and magnetic braking may not explain the observed trend of rotation properties. As we ascend the main sequence the effective temperature, luminosity and mass increase. The more massive stars rotate faster because they have higher initial angular momenta \citep{2013EAS....62..143B}.

\section{Conclusion}
\label{concl}
The rotation frequencies of both normal A and Am/Fm stars were obtained from their frequency spectra using ``hump and spike'' features. Rotation frequency and velocity are found to depend moderately on $\rm T_{eff}$. A good agreement between the rotational velocities and the available $\nu $~sin~$i$ from the literature was observed. However, a comparison of the two velocities would be more feasible if $\nu $~sin~$i$ from high-resolution spectroscopy was available and $i$ was known. The photometric amplitudes were determined and used as proxies for the size of the starspots. The average photometric amplitude of Am/Fm and normal A stars was found to be $\rm \sim 21\pm2 ~ ppm$ and $\rm \sim 19\pm2~ppm$, respectively. After assuming a single, circular and black spot to produce the amplitude of the spike, the average spot radius was found to be $\rm 1.01\,\pm\,0.13$ and $\rm 1.16\,\pm\,0.12$\,$\rm R_E$ for Am/Fm and normal A stars, respectively. These results indicate that spots in normal A and Am/Fm stars are significantly smaller compared to the previously estimated sizes, $\rm 541\pm86\,ppm$ for normal A and $\rm 93\pm25\,ppm$ for Am/Fm stars by \citet{balona13} and \citet{balona15}, respectively. However, based on the limitations discussed in Sect.\,\ref{sss}, our analysis gives the lower limit of the spot sizes.

By fitting the autocorrelation functions of the light-curves with appropriate models, no peaks were observed in the ACFs of some stars, a phenomenon that suggests the absence of repeated patterns in their light-curves, and hence of co-rotating structures (starspots). Since such stars have ``hump and spike'' features in the periodograms, starspots may not be a major requirement to produce spikes in frequency spectra. Based on our $\rm R_{spot}$ values and absence of peaks in the 40\,\% of the ACFs, spots in most normal A and Am/Fm stars could be very weak or actually absent.

The lifetime of the starspots were found to be in the order of a few days. The sizes of the spots in normal A and Am/Fm stars are in general very similar. In comparison with G, K, and M stars, the spots in normal A and Am/Fm stars are short-lived. This is attributed to the much thinner outer convective zone in normal A and Am/Fm than in G, K, and M stars.

In the next paper of the series, we shall present the high-resolution spectroscopic analysis of a number of ``hump and spike'' stars observed from La\,Palma using High-Efficiency and high-Resolution Mercator Echelle Spectrograph (HERMES). In the near future, we also intend to make follow-up studies for K2 and {\it TESS} targets showing ``hump and spike'' features in their frequency spectra.

\section*{Acknowledgement}
The International Science Program (ISP) of Uppsala University financed the study. The part of work presented here is supported by the Belgo-Indian Network for Astronomy \& Astrophysics (BINA), approved by the International Division, Department of Science and Technology (DST, Govt. of India; DST/INT/Belg/P-02) and the Belgian Federal Science Policy Office (BELSPO, Govt. of Belgium; BL/33/IN12). This paper includes data collected by the Kepler mission that were obtained  from  the Mikulski Archive for Space Telescopes (MAST). The authors thank Professor Donald W. Kurtz and Professor Luis A. Balona for the discussions and suggestions regarding this work. OT and EJ acknowledge the hospitality given by ARIES during their research visit to initiate the collaboration between ARIES, India and Mbarara University of Science and Technology, Uganda. We acknowledge the anonymous reviewer for his/her careful reading of our manuscript and  many insightful comments and suggestions which improved the paper.




\bibliographystyle{mnras}
\bibliography{ref} 



\appendix
\section{Tables}

\begin{landscape}
\begin{table}
\caption{List of normal A stars with ``hump and spike'' features in their frequency spectra. We list the rotation frequency ($\rm f_{rot}$), rotational velocity ($\rm V_{rot}$), rotation frequency amplitude ($\rm A_{rot}$), frequency of the first harmonic of the rotation frequency ($\rm f_{har}$) and its amplitude ($\rm A_{har}$) as derived from their frequency spectra, the effective temperature ($\rm T_{eff}$) and surface gravity (log\,g) as reported by \citet{2017ApJS..229...30M}, the luminosity (log(L/$\rm L_{\odot}$)) as derived from Gaia parallaxes \citep{2018yCat.1345....0G}, the stellar radius (R) as estimated from the effective temperature and luminosity, the projected rotational velocity ($\nu $~sin~$i$) as found in the literature \citep{2016A&A...594A..39F} in general; \citet{2017MNRAS.466.3060P} for KIC\,11443271, the spot radius ($\rm R_{spot}$) as estimated from the photometric amplitude, the rotation period ($\rm P_{ACF}$) and decay-time scale of the starspots ($\rm \tau_{DT}$) as estimated from the autocorrelation functions of their light-curves. (*) marks doubtful measurements.}
\label{table1}
\fontsize{7.5}{9.5}\selectfont
\begin{tabular}{lccccccccccccr}
\hline\hline\\
 KIC &$\rm f_{rot}$ &$\rm A_{rot}$ &$\rm f_{har}$ &$\rm A_{har}$ &$\rm T_{eff}$&log\,g&log(L/$\rm L_{\odot}$)&R&$\rm V_{rot}$&$\nu $~sin~$i$&$\rm R_{spot}$ &$\rm P_{ACF}$&$\rm \tau_{DT}$ \\\\
 &$\rm (d^{-1})$&$\rm (mmag)$&$\rm (d^{-1})$&$\rm (mmag)$&(K)&(dex)&(dex)&($\rm R_{\odot}$)&$\rm (km~s^{-1})$&$\rm (km~s^{-1})$&($\rm R_{E}$)&(d)&(d)\\
\hline
 1873552 & 1.21230	& 0.013 $\pm$ 0.002 & 2.40525 & 0.004 $\pm$ 0.002 & 7884 $\pm$ 83 &3.804 $\pm$ 0.152& 1.080 $\pm$ 0.044 & 1.86 $\pm$ 0.06 & 114 $\pm$ 1&$<$ 120 & 0.73 $\pm$ 0.11 & - & 2.923 (*)\\
  2157489 &0.73666 & 0.051 $\pm$ 0.002 & 1.47356 & 0.007 $\pm$ 0.002 & 7645 $\pm$ 270&3.660 $\pm$ 0.275 & 1.034 $\pm$ 0.047 & 1.88 $\pm$ 0.14 & 70 $\pm$ 1 &$<$ 120 & 1.47 $\pm$ 0.12 & - & 1.819 (*)\\
  2158190 & 1.01611	 &0.019 $\pm$ 0.002 & 2.03147 & 0.005 $\pm$ 0.001 & 8204 $\pm$ 294 &3.723 $\pm$ 0.271 & 1.594 $\pm$ 0.048 & 3.11 $\pm$ 0.23 & 156 $\pm$ 2 &$<$ 120 & 1.48 $\pm$ 0.14 & - & 0.699 (*)\\
  3222104 & 1.26946  &0.020 $\pm$ 0.001 & 2.53910 & 0.007 $\pm$ 0.001 & 8457 $\pm$ 318&4.121 $\pm$ 0.142 & 1.475 $\pm$ 0.049 & 2.55 $\pm$ 0.20 & 164 $\pm$ 2 &$<$ 120 & 1.25 $\pm$ 0.12 & - & 3.285 (*)\\
  3337124 &1.10823 & 0.022 $\pm$ 0.001 &2.21620 & 0.004 $\pm$ 0.001 & 7982 $\pm$ 81&3.909 $\pm$ 0.135 & 1.194 $\pm$ 0.046 & 2.07 $\pm$ 0.06 & 116 $\pm$ 1 &$<$ 120 & 1.01 $\pm$ 0.11 & 0.756 $\pm$  0.110& 0.687 $\pm$ 0.108\\
  3634487 & 1.39543 & 0.022 $\pm$ 0.001 & 2.79097& 0.003 $\pm$ 0.001 & 8765 $\pm$ 344&3.552 $\pm$ 0.331&1.676 $\pm$ 0.052 & 2.99 $\pm$ 0.24 & 211 $\pm$ 3 &$<$ 120 &1.53 $\pm$ 0.13 & 0.633 $\pm$  0.112& 0.735 $\pm$ 0.109\\
  3766112 & 2.23214 & 0.011 $\pm$ 0.001 & 4.46765 & 0.005 $\pm$ 0.001 & 9564 $\pm$ 644&3.931 $\pm$ 0.231 & 1.781 $\pm$ 0.054 & 2.84 $\pm$  0.38 & 320 $\pm$ 8 &332 $\pm$ 33 & 1.03 $\pm$ 0.14 & - & 0.655 (*)\\
  3848948 & 1.28684 & 0.020 $\pm$ 0.001 & 2.57368 & 0.002 $\pm$ 0.001 & 8201 $\pm$ 293&3.693 $\pm$ 0.290 & 1.174 $\pm$ 0.043 & 1.92 $\pm$ 0.14 & 125 $\pm$ 2 &- & 0.94 $\pm$ 0.12 &0.736 $\pm$  0.112& 0.726 $\pm$ 0.108\\
  3868032 & 1.67889 & 0.032 $\pm$ 0.002 & 3.35711 & 0.006 $\pm$ 0.002 & 8487 $\pm$ 318&4.042 $\pm$ 0.160 & 1.512 $\pm$ 0.044 & 2.64 $\pm$ 0.20 & 224 $\pm$ 3 &- & 1.63	$\pm$ 0.12 & - & 7.578 (*)\\
  4059089 & 2.62009& 0.051 $\pm$ 0.001 & 5.23999 & 0.012 $\pm$ 0.001 & 9964 $\pm$ 83&3.789 $\pm$ 0.210 & 1.612 $\pm$ 0.047 & 2.15 $\pm$ 0.05 & 285 $\pm$ 2 &280 $\pm$ 72 & 1.68 $\pm$ 0.11 & 0.368 $\pm$  0.100& 0.753 $\pm$ 0.108\\
  4481029 & 1.45328 & 0.004 $\pm$ 0.001 & 2.91331 & 0.001 $\pm$ 0.001 & 9009 $\pm$ 81&3.878 $\pm$ 0.138 & 1.262 $\pm$ 0.050 & 1.76 $\pm$ 0.05 & 129 $\pm$ 1 &$<$ 120 & 0.38 $\pm$ 0.11 &0.633 $\pm$  0.105&  0.719 $\pm$ 0.108\\
  4488313 & 1.19864 & 0.019 $\pm$ 0.001 & 2.39748 & 0.007 $\pm$ 0.001 & 8374 $\pm$ 78&3.766 $\pm$ 0.125 & 1.588 $\pm$ 0.048 & 2.96 $\pm$ 0.07 & 180 $\pm$ 1 &$<$ 120 & 1.41 $\pm$ 0.11 & - & 3.685 (*)\\
  4567097 & 1.04853 & 0.006 $\pm$ 0.001 & 2.09765 & 0.004 $\pm$ 0.001 & 9849 $\pm$ 359&4.205 $\pm$ 0.195 & 2.313 $\pm$ 0.046 & 4.93 $\pm$ 0.36 & 262 $\pm$ 2 &- & 1.32 $\pm$ 0.10 & 2.861 $\pm$  0.102& 4.72 $\pm$  0.161\\
  4572373 & 1.47522 & 0.021 $\pm$ 0.001 & 2.95112 & 0.005 $\pm$ 0.001 & 9100 $\pm$ 350&3.681 $\pm$ 0.290 & 1.541 $\pm$ 0.046 & 2.38 $\pm$ 0.19 & 177 $\pm$ 2&- & 1.19 $\pm$ 0.10 & 0.674 $\pm$  0.116& 2.26 $\pm$  0.128\\
  4661914 & 0.97690 & 0.019 $\pm$ 0.001 & 1.95360 & 0.008 $\pm$ 0.001 & 7652 $\pm$ 281&3.499 $\pm$ 0.290 & 0.975 $\pm$ 0.042 & 1.75 $\pm$ 0.13 & 87 $\pm$ 1 &- & 0.83 $\pm$ 0.10 & 0.981 $\pm$  0.102& 0.665 $\pm$ 0.1089\\
  4669058 & 1.09085 & 0.019 $\pm$ 0.012 & 2.18203 & 0.003 $\pm$ 0.001 & 8080 $\pm$ 281&3.859 $\pm$ 0.210 & 1.054 $\pm$ 0.046 & 1.72 $\pm$ 0.12 & 95 $\pm$ 1 &- & 0.82 $\pm$ 0.15 & - & 1.609 (*)\\
  4818496 & 1.62697 & 0.014 $\pm$ 0.002 & 3.25483 & 0.003 $\pm$ 0.001 & 8857 $\pm$ 322&3.687 $\pm$ 0.330 & 1.529 $\pm$ 0.042 & 2.48 $\pm$ 0.18 & 204 $\pm$ 3 & - & 1.01 $\pm$ 0.13 & 0.593 $\pm$  0.264&  0.774 $\pm$ 0.109\\
  4856799 & 1.50678 & 0.022 $\pm$ 0.001 & 3.01300 & 0.002 $\pm$ 0.001 & 9520 $\pm$ 592&3.909 $\pm$ 0.225 & 1.686 $\pm$ 0.047&2.56 $\pm$ 0.32&196 $\pm$ 4 &$<$ 120 & 1.31 $\pm$ 0.14 & 0.531 $\pm$  0.123& 0.645 $\pm$ 0.107\\
  4857593 & 1.80324 & 0.020 $\pm$ 0.001 & 3.60407 & 0.004 $\pm$ 0.001 & 8279 $\pm$ 78&4.116 $\pm$ 0.126 & 1.523 $\pm$ 0.053 & 2.81 $\pm$ 0.07 & 257 $\pm$ 4 &175 $\pm$ 89& 1.37 $\pm$ 0.11 & 0.450 $\pm$  0.116& 0.618 $\pm$ 0.106\\
  4863277 & 2.07037 & 0.006 $\pm$ 0.002 & 4.14581 & 0.002 $\pm$ 0.001 & 9858 $\pm$ 355&4.205 $\pm$ 0.200 & 1.682 $\pm$ 0.056 & 2.38 $\pm$ 0.18 & 250 $\pm$ 4 &207 $\pm$ 78 & 0.64 $\pm$ 0.13 & -& 1.197 (*)\\
  4935271 & 0.54163 & 0.034 $\pm$ 0.002 & 1.08881 & 0.006 $\pm$ 0.002 & 7859 $\pm$283&4.080 $\pm$ 0.145 & 1.408 $\pm$ 0.055 & 2.73 $\pm$ 0.20 & 75 $\pm$ 1 &- & 1.74 $\pm$ 0.13 & - & 2.799 (*)\\
  4944828 & 1.09952 & 0.020 $\pm$ 0.001 & 2.19890 & 0.007 $\pm$ 0.001&7983 $\pm$ 80&3.730 $\pm$ 0.157 & 1.382 $\pm$ 0.044 & 2.57 $\pm$ 0.07 & 143 $\pm$ 1 &$<$ 120 & 1.26 $\pm$ 0.11 & 0.920 $\pm$ 0.104&3.629 $\pm$ 0.136\\
  4953140 & 2.74956 & 0.022 $\pm$ 0.002 & 5.49926 & 0.007 $\pm$ 0.002 & 8855 $\pm$ 340&3.826 $\pm$ 0.250 & 1.408 $\pm$ 0.051 & 2.15 $\pm$ 0.17 & 300 $\pm$ 6 &320 $\pm$ 39& 1.10 $\pm$ 0.13 & - & 0.647 (*)\\
  5039203 & 1.42323 & 0.017 $\pm$ 0.001 & 2.84605 & 0.004 $\pm$ 0.001 & 8204 $\pm$ 293&4.020 $\pm$ 0.160 & 1.298 $\pm$ 0.047 & 2.21 $\pm$ 0.16 & 159 $\pm$ 2 &- & 1.00 $\pm$ 0.12 & - &0.604 (*)\\
  5106287 & 1.89345 & 0.015 $\pm$ 0.002 & 3.78851 & 0.003 $\pm$ 0.002 & 7804 $\pm$ 286&3.750 $\pm$ 0.228 & 1.433 $\pm$ 0.048 & 2.85 $\pm$ 0.21 & 273 $\pm$ 4&- & 1.21 $\pm$ 0.13 & -&0.619 (*)\\
  5129737 & 1.17568 & 0.013 $\pm$ 0.001 & 2.35980 & 0.004 $\pm$ 0.001 & 8368 $\pm$ 298&3.748 $\pm$ 0.285 & 1.809 $\pm$ 0.062 & 3.83 $\pm$ 0.28 & 228 $\pm$ 2 &- & 1.51 $\pm$ 0.13 & -&0.954 (*)\\
  5213792 & 2.08696 & 0.009 $\pm$ 0.001 & 4.21858 & 0.002 $\pm$ 0.001 & 9769 $\pm$ 342&4.213 $\pm$ 0.195 & 1.791 $\pm$ 0.053 & 2.75 $\pm$ 0.20 & 290 $\pm$ 4&- & 0.90 $\pm$ 0.12 & -&2.014 (*)\\
  5273195 & 1.58529 & 0.005 $\pm$ 0.001 & 3.17646 & 0.001 $\pm$ 0.001 & 8732 $\pm$ 80&3.806 $\pm$ 0.165 & 1.311 $\pm$ 0.052 & 1.98 $\pm$ 0.05 & 159 $\pm$ 1 &$<$ 120 & 0.48 $\pm$ 0.11 & -&5.268 (*)\\
  5288746 & 1.45866 & 0.005 $\pm$ 0.001 & 2.91718 & 0.002 $\pm$ 0.001 & 9153 $\pm$ 358&4.106 $\pm$ 0.160 & 1.248 $\pm$ 0.045 & 1.68 $\pm$ 0.13 & 124 $\pm$ 2&- & 0.41 $\pm$ 0.11 & - & 1.639 (*)\\
  5387719 & 1.53768 & 0.020 $\pm$ 0.001 & 3.07791 & 0.003 $\pm$ 0.001 & 9103 $\pm$ 340&3.857 $\pm$ 0.240 & 1.799 $\pm$ 0.045 &3.20 $\pm$ 0.24&249 $\pm$ 3&$<$ 120 & 1.56 $\pm$ 0.13 & 0.633 $\pm$  0.131& 0.862 $\pm$ 0.110\\
  5392475 & 0.55034 & 0.011 $\pm$ 0.001 & 1.15307 & 0.004 $\pm$ 0.001 & 9211 $\pm$ 357&4.077 $\pm$ 0.180 & 1.862 $\pm$ 0.057 & 3.35 $\pm$ 0.27 & 94 $\pm$ 1 &- & 1.21 $\pm$ 0.13 & - & 0.864 (*)\\
  5395418 & 1.67162 & 0.033 $\pm$ 0.001 & 3.34623 & 0.005 $\pm$ 0.001 & 9164 $\pm$ 333&3.865 $\pm$ 0.245 & 1.967 $\pm$ 0.044 & 3.82 $\pm$ 0.28 & 323 $\pm$ 4&- & 2.40 $\pm$ 0.13 & 0.593 $\pm$  0.106& 5.188 $\pm$ 0.154\\
  5456027 & 1.24204 & 0.015 $\pm$0.001 & 2.47967 & 0.008 $\pm$ 0.001 & 8346 $\pm$ 280&3.786 $\pm$ 0.257 & 1.417 $\pm$ 0.046 & 2.45 $\pm$ 0.17 & 154 $\pm$ 2&- & 1.04 $\pm$ 0.12 & 0.817 $\pm$  0.103& 4.070 $\pm$ 0.171\\
  5524045 & 1.81517 & 0.023 $\pm$ 0.001 & 3.63340 & 0.004 $\pm$ 0.001 & 9500 $\pm$ 82&3.869 $\pm$ 0.155&1.604 $\pm$ 0.043 & 2.34 $\pm$ 0.05 & 215 $\pm$ 2&201 $\pm$ 58 & 1.23 $\pm$ 0.10 &0.552 $\pm$  0.309 &  2.320 $\pm$  0.125\\ 
  5545321 & 0.86908 & 0.015 $\pm$ 0.003 & 1.73809 & 0.008 $\pm$ 0.001 & 7807 $\pm$ 596&3.635 $\pm$ 0.290&1.043 $\pm$ 0.043 & 1.82 $\pm$ 0.28 & 80 $\pm$ 2&$<$ 120 & 0.77 $\pm$ 0.15 &1.042 $\pm$ 0.116&0.803 $\pm$ 0.109\\
  5566579 & 1.53465 & 0.025 $\pm$ 0.009 & 3.07023 & 0.003 $\pm$ 0.001 & 9189 $\pm$ 344&4.198 $\pm$ 0.150& 1.416 $\pm$ 0.046 & 2.02 $\pm$ 0.15 & 157 $\pm$ 2&- & 1.10 $\pm$ 0.15 & -& 1.522 (*)\\
  5641490 & 0.97187 & 0.026 $\pm$0.001 & 1.94347 & 0.004 $\pm$ 0.001 & 9430 $\pm$ 83&3.826 $\pm$ 0.145 & 1.644 $\pm$ 0.048 & 2.49 $\pm$ 0.06 & 123 $\pm$ 1&$<$ 120 & 1.39 $\pm$ 0.11 & - & 2.106 (*)\\
  5650229 & 2.16673 & 0.022 $\pm$ 0.003 & 4.33197 & 0.004 $\pm$ 0.001 & 8763 $\pm$ 344&3.870 $\pm$ 0.240 & 1.648 $\pm$ 0.044 & 2.90 $\pm$ 0.23 & 318 $\pm$ 5&213 $\pm$ 61 & 1.49 $\pm$ 0.14 & 0.429 $\pm$  0.101& 0.704 $\pm$ 0.111\\
  5730714 & 0.50930 & 0.022 $\pm$ 0.001 & 1.01883 & 0.008 $\pm$ 0.001 & 9765 $\pm$ 376&3.646 $\pm$ 0.300 & 2.215 $\pm$ 0.057 & 4.48 $\pm$ 0.35 & 115 $\pm$ 1 &- & 2.29 $\pm$ 0.14 & 1.900  $\pm$  0.124& 0.725 $\pm$  0.112\\
  5897685 & 1.63668 & 0.016 $\pm$ 0.001 & 3.27303 & 0.002 $\pm$ 0.001 & 9081 $\pm$ 358&4.130 $\pm$ 0.158 & 1.373 $\pm$ 0.046 & 1.97 $\pm$ 0.16 & 163 $\pm$ 3&- & 0.86 $\pm$ 0.12 & - & 0.633 (*)\\
  5903499 & 2.48245 & 0.007 $\pm$ 0.001 & 4.95973 & 0.002 $\pm$ 0.001 & 9282 $\pm$ 357&4.183 $\pm$ 0.165 & 1.363 $\pm$ 0.053 & 1.86 $\pm$ 0.15 & 234 $\pm$ 4&- & 0.54 $\pm$ 0.12 & 0.347 $\pm$  0.175 & 0.701 $\pm$ 0.107\\
\hline
 \end{tabular}
\end{table}
\end{landscape}

\begin{landscape}
\begin{table}
\contcaption{ }
\fontsize{7.5}{9.5}\selectfont
\begin{tabular}{lccccccccccccr}
\hline\hline\\
 KIC &$\rm f_{rot}$ &$\rm A_{rot}$ &$\rm f_{har}$ &$\rm A_{har}$ &$\rm T_{eff}$&log\,g&log(L/$\rm L_{\odot}$)&R&$\rm V_{rot}$&$\nu $~sin~$i$&$\rm R_{spot}$ &$\rm P_{ACF}$&$\rm \tau_{DT}$ \\\\
 &$\rm (d^{-1})$&$\rm (mmag)$&$\rm (d^{-1})$&$\rm (mmag)$&(K)&(dex)&(dex)&($\rm R_{\odot}$)&$\rm (km~s^{-1})$&$\rm (km~s^{-1})$&($\rm R_{E}$)&(d)&(d)\\
\hline
  5905878 & 1.25277 & 0.009 $\pm$ 0.001 & 2.50541 & 0.003 $\pm$ 0.001 & 8481 $\pm$ 302&4.050 $\pm$ 0.156 & 1.161 $\pm$ 0.043 & 1.77 $\pm$ 0.13 & 112 $\pm$ 2 &- & 0.58 $\pm$ 0.11 & - & 0.213 (*)\\
  5978118 & 2.57045 & 0.041 $\pm$ 0.001 & 5.14291 & 0.006 $\pm$ 0.001 & 9345 $\pm$ 351&3.955 $\pm$ 0.209 & 1.602 $\pm$ 0.051 & 2.42 $\pm$ 0.19 & 314 $\pm$ 5 &312 $\pm$ 37 & 1.69 $\pm$ 0.12 & 0.368 $\pm$  0.186 &  0.643 $\pm$ 0.106\\
  5980337 & 1.29218 & 0.008 $\pm$ 0.001 & 2.58307 & 0.003 $\pm$ 0.001 & 7825 $\pm$ 79&3.984 $\pm$ 0.114 & -0.044 $\pm$ 0.114 & 0.52 $\pm$ 0.68 & 34 $\pm$ 5 &$<$ 120& 0.16 $\pm$ 0.07 & - & 0.560 (*)\\
  6062984 & 1.81630 & 0.002 $\pm$ 0.001 & 3.64138 & 0.001 $\pm$ 0.001 & 9348 $\pm$ 376&4.009 $\pm$ 0.200 & 1.572 $\pm$ 0.045 & 2.33 $\pm$ 0.19 & 215 $\pm$ 3 &184 $\pm$ 53& 0.36 $\pm$ 0.10 & 0.450 $\pm$  0.112 & 0.663 $\pm$ 0.107\\
  6114539 & 1.06152  & 0.012 $\pm$ 0.001 & 2.12341 & 0.003 $\pm$ 0.001 & 7785 $\pm$ 283&4.049 $\pm$ 0.150 & 1.058 $\pm$ 0.043 & 1.86 $\pm$ 0.14 & 100 $\pm$ 2 &- & 0.70 $\pm$ 0.12 & - & 1.783 (*)\\
   6131612 & 1.16971 & 0.006 $\pm$ 0.001 & 2.33885 & 0.001 $\pm$ 0.001 & 9037 $\pm$ 84&3.755 $\pm$ 0.185 & 1.651 $\pm$ 0.053 & 2.73 $\pm$ 0.07 & 162 $\pm$ 1 &$<$ 120 & 0.73 $\pm$ 0.11 & - & 3.100 (*)\\
  6147122 & 0.95630 & 0.007 $\pm$ 0.001 & 1.91426 & 0.007 $\pm$ 0.001 & 7769 $\pm$ 270&4.080 $\pm$ 0.145 & 0.980 $\pm$ 0.044 & 1.71 $\pm$ 0.12 & 83 $\pm$ 1 &-& 0.49 $\pm$ 0.11 & - & 1.073 (*)\\
  6192566 & 0.75287 & 0.029 $\pm$ 0.008 & 1.50679 & 0.009 $\pm$ 0.001 &7689 $\pm$ 270&3.861 $\pm$ 0.210&1.655 $\pm$ 0.043 & 3.80 $\pm$ 0.27 & 145 $\pm$ 2 &$<$ 120 & 2.23 $\pm$ 0.18 & 1.246 $\pm$  0.103 & 0.760 $\pm$ 0.114\\
  6222381 & 1.22897 & 0.024 $\pm$ 0.001 & 2.45817 & 0.007 $\pm$ 0.001& 8597 $\pm$ 78&3.761 $\pm$ 0.152&1.526 $\pm$ 0.045 & 2.62 $\pm$ 0.06 & 163 $\pm$ 1 &$<$ 120 & 1.40 $\pm$ 0.11 & 0.817 $\pm$  0.104 & 3.623 $\pm$ 0.137\\
  6266219 & 0.89751 & 0.017 $\pm$ 0.001 & 1.79499 & 0.003 $\pm$ 0.001 & 7928 $\pm$ 281&4.048 $\pm$ 0.156 & 1.177 $\pm$ 0.043 & 2.06 $\pm$ 0.15 & 94 $\pm$ 1 &-& 0.93 $\pm$ 0.10 & 1.083 $\pm$  0.117 &  0.892 $\pm$ 0.111\\
  6450107 & 1.53317 & 0.008 $\pm$ 0.001 & 3.06111 & 0.001 $\pm$ 0.001 & 9800 $\pm$ 362&4.053 $\pm$ 0.190 & 1.878 $\pm$ 0.042 & 3.02 $\pm$ 0.23 & 234 $\pm$ 3 &- & 0.93 $\pm$ 0.10 & 0.633  $\pm$ 0.124 & 0.865 $\pm$ 0.110\\
  6468647 & 1.74231 & 0.016 $\pm$ 0.001 & 3.48446 & 0.003 $\pm$ 0.001 & 9073 $\pm$ 358&4.041 $\pm$ 0.169 & 1.328 $\pm$ 0.047 & 1.87 $\pm$ 0.15 & 165 $\pm$ 3&- & 0.82 $\pm$ 0.12 & - & 0.557 (*)\\
  6530137 & 1.48994 & 0.015 $\pm$ 0.001 & 2.98002 & 0.012 $\pm$ 0.001 & 8650 $\pm$ 301&4.140 $\pm$ 0.140 & 1.368 $\pm$ 0.048 & 2.15 $\pm$ 0.15 & 163 $\pm$ 2&- & 0.91 $\pm$ 0.12 & 0.593 $\pm$  0.103 & 0.662 $\pm$ 0.107\\
  6794857 & 0.94616 & 0.021 $\pm$ 0.001 & 1.89270 & 0.011 $\pm$ 0.001 & 8344 $\pm$ 298&3.930 $\pm$ 0.205 & 1.523 $\pm$ 0.055 & 2.77 $\pm$ 0.20 & 132 $\pm$ 2 &$<$ 120 & 1.39 $\pm$ 0.12 & - & 0.588 (*)\\
  6871866 & 1.04369 & 0.018 $\pm$ 0.001 & 2.10348 & 0.013 $\pm$ 0.001 & 7781 $\pm$ 270&3.966 $\pm$ 0.180 & 1.098 $\pm$ 0.045 & 1.95 $\pm$ 0.14 & 103 $\pm$ 2 &- & 0.90 $\pm$ 0.12 & - & 0.130 (*)\\
  6974705 & 1.61395 & 0.033 $\pm$ 0.001 & 3.22916 & 0.007 $\pm$ 0.001 & 8769 $\pm$ 334&3.856 $\pm$ 0.245 & 1.568 $\pm$ 0.046 & 2.64 $\pm$ 0.20 & 215 $\pm$ 3&- & 1.66 $\pm$ 0.12 & 0.613 $\pm$  0.214 & 0.912 $\pm$ 0.110\\
  7129454 & 1.93681 & 0.019 $\pm$ 0.001 & 3.87528 & 0.003 $\pm$ 0.001 & 9261 $\pm$ 357 &4.140 $\pm$ 0.160 & 1.606 $\pm$ 0.052 & 2.47 $\pm$ 0.20 & 242 $\pm$ 4 &249 $\pm$ 71& 1.18 $\pm$ 0.12 & 0.511 $\pm$  0.115 & 0.813 $\pm$ 0.109\\
  7375997 & 1.21081 & 0.043 $\pm$0.001 & 2.42148 & 0.008 $\pm$ 0.001 & 8554 $\pm$ 305&3.780 $\pm$ 0.257&1.395 $\pm$ 0.048 & 2.27 $\pm$ 0.17 & 139 $\pm$ 2 &$<$ 120 & 1.63 $\pm$ 0.10 & 0.817 $\pm$  0.105 & 1.739 $\pm$ 0.123\\
  7530366 & 1.33506 & 0.021 $\pm$ 0.001 & 2.66863 & 0.003 $\pm$ 0.001 & 9700 $\pm$ 388&3.567 $\pm$ 0.325 & 1.922 $\pm$ 0.043 & 3.24 $\pm$ 0.26 & 218 $\pm$ 3&- & 1.62 $\pm$ 0.10 & 0.756 $\pm$  0.104 & 3.380 $\pm$ 0.140\\
  7622596 & 1.67546 & 0.008 $\pm$ 0.001 & 3.35245 & 0.001 $\pm$ 0.001& 10007 $\pm$ 335&3.572 $\pm$ 0.265 & 2.002 $\pm$ 0.052 & 3.34 $\pm$ 0.23 & 283 $\pm$ 3&156 $\pm$ 57& 1.03 $\pm$ 0.12 & - & 7.446 (*)\\
  7667560 & 1.54370 & 0.011 $\pm$ 0.001 & 3.08692 & 0.003 $\pm$ 0.001 & 8590 $\pm$ 323&4.080 $\pm$ 0.150 & 1.188 $\pm$ 0.042 & 1.78 $\pm$ 0.14 & 139 $\pm$ 2&- & 0.64 $\pm$ 0.10 & 0.633 $\pm$  0.119 & 0.736 $\pm$ 0.109\\
  7939835 & 0.63143 & 0.007 $\pm$ 0.001 & 1.26283 & 0.002 $\pm$ 0.001& 7786 $\pm$ 286&3.819 $\pm$ 0.220 & 1.278 $\pm$ 0.044 & 2.40 $\pm$ 0.18 & 77 $\pm$ 1&$<$ 120 & 0.69 $\pm$ 0.12 & - & 4.110 (*)\\
  7959579 & 1.20814 & 0.009 $\pm$ 0.001 & 2.41713 & 0.008 $\pm$ 0.001 & 7696 $\pm$ 268&4.043 $\pm$ 0.160&1.010 $\pm$ 0.042 & 1.80 $\pm$ 0.13 &110 $\pm$ 2 &$<$ 120& 0.59 $\pm$ 0.10 &0.797 $\pm$  0.103 & 0.696 $\pm$ 0.107\\
  8043385 & 1.85082 & 0.017 $\pm$ 0.001 & 3.70975 & 0.003 $\pm$ 0.001 & 9818 $\pm$ 359&4.160 $\pm$ 0.200 & 2.475 $\pm$ 0.065 & 5.99 $\pm$ 0.45 & 547 $\pm$ 6 &250 $\pm$ 73 & 2.70 $\pm$ 0.15 & 0.511 $\pm$ 0.107 & 0.637 $\pm$ 0.106\\
  8121280 & 1.34050 & 0.028 $\pm$ 0.001 & 2.68096 & 0.008 $\pm$ 0.001 & 8256 $\pm$ 274&3.764 $\pm$ 0.268 & 1.291 $\pm$ 0.048 & 2.16 $\pm$ 0.15 & 147 $\pm$ 2 &- & 1.25 $\pm$ 0.12 & - & 1.734 (*)\\
  8191793 & 1.64003 & 0.023 $\pm$ 0.002 & 3.28023 & 0.008 $\pm$ 0.001 & 8066 $\pm$ 283&4.002 $\pm$ 0.165 & 1.526 $\pm$ 0.061 & 2.97 $\pm$ 0.22 & 259 $\pm$ 3 &139 $\pm$ 102 & 1.56 $\pm$ 0.13 & - & 2.949 (*)\\
  8330169 & 1.23067 & 0.027 $\pm$ 0.007 & 2.46198 & 0.006 $\pm$ 0.002 & 8075 $\pm$ 283&3.741 $\pm$ 0.262 & 1.642 $\pm$ 0.048 & 3.39 $\pm$ 0.24 & 211 $\pm$ 2 &159 $\pm$ 89 & 1.92 $\pm$ 0.17 & 0.776 $\pm$  0.108 & 0.682 $\pm$ 0.107\\
  8330740 & 1.14406 & 0.032 $\pm$ 0.001 & 2.48480 & 0.006 $\pm$ 0.001 & 7951 $\pm$ 276&4.048 $\pm$ 0.156 & 1.346 $\pm$ 0.045 & 2.49 $\pm$ 0.18 & 144 $\pm$ 2 &- & 1.54 $\pm$ 0.12 & - & 1.519 (*)\\
  8385850 & 1.30554 & 0.011 $\pm$ 0.001 & 2.68508 & 0.001 $\pm$0.001 & 8291 $\pm$ 290&3.758 $\pm$ 0.289 & 1.488 $\pm$ 0.053 & 2.69 $\pm$ 0.19 & 178 $\pm$ 2 &- & 0.97 $\pm$ 0.12 & - & 0.603 (*)\\
  8391713 & 1.70470 & 0.009 $\pm$ 0.001 & 3.42719 & 0.001 $\pm$ 0.001 & 8918 $\pm$ 354&3.813 $\pm$ 0.278 & 1.811 $\pm$ 0.051 & 3.38 $\pm$ 0.27 & 291 $\pm$ 4 &134 $\pm$ 100 & 1.11 $\pm$ 0.13 & - & 0.577 (*)\\
  8396240 & 0.98459 & 0.020 $\pm$ 0.002 & 1.96914 & 0.005 $\pm$ 0.001 & 7793 $\pm$ 278&4.028 $\pm$ 0.165 & 1.208 $\pm$ 0.045 & 2.21 $\pm$ 0.16 & 110 $\pm$ 2 &- & 1.08 $\pm$ 0.13 & - & (*)\\
  8396309 & 1.09041 & 0.023 $\pm$ 0.002 & 2.18707 & 0.004 $\pm$ 0.002 & 8101 $\pm$ 297&3.725 $\pm$ 0.294 & 1.687 $\pm$ 0.053 & 3.55 $\pm$ 0.27 & 196 $\pm$ 2 &154 $\pm$ 104 & 1.86 $\pm$ 0.14 & - & 1.954 (*)\\
  8396872 & 1.43485 & 0.027 $\pm$ 0.001 & 2.86954 & 0.004 $\pm$ 0.002 & 8002 $\pm$ 280&3.713 $\pm$ 0.270 & 1.918 $\pm$ 0.051 & 4.74 $\pm$ 0.34 & 344 $\pm$ 4 &$<$ 120 & 2.69 $\pm$ 0.14 & 0.674 $\pm$  0.133 & 0.751 $\pm$ 0.108\\
  8453527 & 1.22669 & 0.036 $\pm$ 0.001 & 2.45335 & 0.008 $\pm$ 0.001 & 8141 $\pm$ 285&3.725 $\pm$ 0.294 & 1.538 $\pm$ 0.044 & 2.96 $\pm$ 0.21 & 184 $\pm$ 2 &$<$ 120 & 1.94 $\pm$ 0.12 & 0.776 $\pm$  0.104 & 0.764 $\pm$ 0.109\\
  8524728 & 2.45960 & 0.022 $\pm$ 0.001 & 4.92597 & 0.006 $\pm$ 0.001 & 8930 $\pm$ 356&3.822 $\pm$ 0.265 & 1.633 $\pm$ 0.049 & 2.74 $\pm$ 0.22 & 342 $\pm$ 6 &- & 1.40 $\pm$ 0.12 & 0.347 $\pm$  0.102 & 0.678 $\pm$ 0.106\\
  8564695 & 1.17960 & 0.010 $\pm$ 0.001 & 2.35883 & 0.005 $\pm$ 0.001 & 7984 $\pm$ 281&3.996 $\pm$ 0.178 & 1.185 $\pm$ 0.049 & 2.05 $\pm$ 0.15 & 122 $\pm$ 2 &$<$ 120 & 0.71 $\pm$ 0.12 & - & 4.484 (*)\\
  8570955 & 1.64329 & 0.005 $\pm$ 0.001 & 3.28764 & 0.002 $\pm$ 0.001 & 8126 $\pm$ 283&4.155 $\pm$ 0.135 & 1.102 $\pm$ 0.048 & 1.80 $\pm$ 0.13 & 149 $\pm$ 2 &- & 0.44 $\pm$ 0.11 & - & 1.725 (*)\\
  8581557 & 2.07836 & 0.017 $\pm$ 0.001 & 4.15866 & 0.003 $\pm$ 0.001 & 8255 $\pm$ 297&4.084 $\pm$ 0.145 & 1.323 $\pm$ 0.047 & 2.25 $\pm$ 0.17 & 236 $\pm$ 4 &197 $\pm$ 60 & 1.01 $\pm$ 0.12 & 0.450 $\pm$ 0.397 & 0.654 $\pm$ 0.106\\
  8586760 & 1.05307 & 0.035 $\pm$ 0.001 & 2.10546 & 0.006 $\pm$ 0.001 & 7785 $\pm$ 285&3.702 $\pm$ 0.270 & 1.802 $\pm$ 0.047 & 4.38 $\pm$ 0.33 & 233 $\pm$ 2 &$<$ 120 & 2.83 $\pm$ 0.14 & 0.940 $\pm$ 0.102 & 0.928 $\pm$ 0.111\\
  8884276 & 2.29528 & 0.029 $\pm$ 0.001 & 4.59533 & 0.011 $\pm$ 0.001 & 9377 $\pm$ 357&3.926 $\pm$ 0.230 & 1.637 $\pm$ 0.057 & 2.50 $\pm$ 0.20 & 290 $\pm$ 5 &- & 1.47 $\pm$ 0.12 & 0.409 $\pm$  0.170 & 0.644 $\pm$ 0.107\\
  9044567 & 0.71195 & 0.010 $\pm$ 0.002 & 1.40316 & 0.003 $\pm$ 0.001 & 8049 $\pm$ 281&3.717 $\pm$ 0.283 & 1.944 $\pm$ 0.077 & 4.83 $\pm$ 0.35 & 174 $\pm$ 2 &$<$ 120 & 1.67 $\pm$ 0.15 & - & 1.742 (*)\\
  9099944 & 1.01448 & 0.018 $\pm$ 0.001 & 2.02883 & 0.004 $\pm$ 0.001 & 8190 $\pm$ 299&3.722 $\pm$ 0.271 & 1.345 $\pm$ 0.049 & 2.34 $\pm$ 0.18 & 120 $\pm$ 2 &- & 1.08 $\pm$ 0.12 & - & 3.217 (*)\\
  9181642 & 1.50649 & 0.026 $\pm$ 0.001 & 3.01331 & 0.004 $\pm$ 0.001 & 7950 $\pm$ 281&4.001 $\pm$ 0.170 & 1.280 $\pm$ 0.045 & 2.31 $\pm$ 0.17 & 176 $\pm$ 2 &$<$ 120 & 1.29 $\pm$ 0.12 & - & 1.634 (*)\\
  9222948 & 0.77452 & 0.256 $\pm$ 0.002 & 1.54815 & 0.071 $\pm$ 0.002 & 8775 $\pm$ 334&3.843 $\pm$ 0.255 & 1.234 $\pm$ 0.046 & 1.79 $\pm$ 0.14 & 71 $\pm$ 1 &$<$ 120 & 3.13 $\pm$ 0.12 & 1.287 $\pm$ 0.109 & 7.371 $\pm$ 0.162\\
\hline
 \end{tabular}
\end{table}
\end{landscape}

\begin{landscape}
\begin{table}
\contcaption{ }
 \fontsize{7.5}{9.5}\selectfont
\begin{tabular}{lccccccccccccr}
\hline\hline\\
 KIC &$\rm f_{rot}$ &$\rm A_{rot}$ &$\rm f_{har}$ &$\rm A_{har}$ &$\rm T_{eff}$&log\,g&log(L/$\rm L_{\odot}$)&R&$\rm V_{rot}$&$\nu $~sin~$i$&$\rm R_{spot}$ &$\rm P_{ACF}$&$\rm \tau_{DT}$ \\\\
 &$\rm (d^{-1})$&$\rm (mmag)$&$\rm (d^{-1})$&$\rm (mmag)$&(K)&(dex)&(dex)&($\rm R_{\odot}$)&$\rm (km~s^{-1})$&$\rm (km~s^{-1})$&($\rm R_{E}$)&(d)&(d)\\
 \hline
  9299980 & 2.26102 & 0.028 $\pm$ 0.002 & 4.52171 & 0.005 $\pm$ 0.002 & 8399 $\pm$ 301&3.755 $\pm$ 0.280 & 1.257 $\pm$ 0.046 & 2.01 $\pm$ 0.15 & 230 $\pm$ 4 &234 $\pm$ 73 & 1.16 $\pm$ 0.12 & 0.409 $\pm$ 0.179 & 0.672 $\pm$ 0.107\\
  9304955 & 1.18183 & 0.037 $\pm$ 0.001 & 2.36407 & 0.007 $\pm$ 0.001 & 8086 $\pm$ 283&4.008 $\pm$ 0.165& 1.546 $\pm$ 0.050 & 3.03 $\pm$ 0.22 & 181 $\pm$ 2 &- & 2.01 $\pm$ 0.10 & - & 1.908 (*)\\
  9396171 & 2.22723 & 0.051 $\pm$ 0.002 & 4.45429 & 0.007 $\pm$ 0.002 & 9563 $\pm$369&4.058 $\pm$ 0.205& 1.708 $\pm$ 0.053 & 2.61 $\pm$ 0.20 & 294 $\pm$ 5 &299 $\pm$ 47& 2.02 $\pm$ 0.13 & 0.450 $\pm$ 0.102 &3.546 $\pm$ 0.141\\
  9418202 & 0.55205 & 0.035 $\pm$ 0.001 & 1.11474 & 0.006 $\pm$ 0.001 & 7763 $\pm$ 285&3.683 $\pm$ 0.270 & 1.444 $\pm$ 0.050 & 2.92 $\pm$ 0.22 & 82 $\pm$ 1&- & 1.89 $\pm$ 0.12 & - & 1.685 (*)\\
  9428798 & 2.28352 & 0.016 $\pm$ 0.001 & 4.56549 & 0.005 $\pm$ 0.001 & 8049 $\pm$ 294&4.004 $\pm$ 0.157 & 1.398 $\pm$ 0.049 & 2.58 $\pm$ 0.19 & 298 $\pm$ 5 &228 $\pm$ 65& 1.13 $\pm$ 0.10 & 0.347 $\pm$  0.102 & 0.711 $\pm$ 0.107\\
  9453452 & 1.64741 & 0.034 $\pm$ 0.001 & 3.29574 & 0.005 $\pm$ 0.001 & 8664 $\pm$ 316&3.901 $\pm$ 0.225 & 1.437 $\pm$ 0.044 & 2.33 $\pm$ 0.17 & 194 $\pm$ 3 &$<$ 120 & 1.48 $\pm$ 0.12 & 0.593 $\pm$  0.103 & 1.016 $\pm$ 0.112\\
  9519698 & 2.44243 & 0.009 $\pm$ 0.001 & 4.89319 & 0.001 $\pm$ 0.001 & 9859 $\pm$ 345&4.205 $\pm$ 0.200& 1.806 $\pm$ 0.048 & 2.75 $\pm$ 0.20 & 340 $\pm$ 5 &$<$ 120 & 0.90 $\pm$ 0.12 & 0.388 $\pm$ 0.115 & 0.662 $\pm$ 0.107\\
  9532445 & 1.06369 & 0.019 $\pm$ 0.001 &2.12729 & 0.004 $\pm$ 0.001 & 7841 $\pm$ 275&3.958 $\pm$ 0.178 & 1.190 $\pm$ 0.048 & 2.14 $\pm$ 0.16 & 115 $\pm$ 2&- & 1.02 $\pm$ 0.12 & 0.797 $\pm$ 0.107 & 0.667 $\pm$ 0.109\\
  9593997 & 2.16319 & 0.042 $\pm$ 0.001 & 4.32566 & 0.053 $\pm$ 0.001 & 7759 $\pm$ 275&3.613 $\pm$ 0.310 & 1.105 $\pm$ 0.048 & 1.98 $\pm$ 0.15 & 216 $\pm$ 3 &147 $\pm$ 97 & 1.40 $\pm$ 0.12 & 0.450 $\pm$ 0.119 & 0.683 $\pm$ 0.108\\
  9596469 & 1.58355 & 0.040 $\pm$ 0.001 & 3.16867 & 0.029 $\pm$ 0.001 & 7687 $\pm$ 80&3.920 $\pm$ 0.120 & 1.121 $\pm$ 0.043 & 2.05 $\pm$ 0.06 & 164 $\pm$ 1 &$<$ 120 & 1.42 $\pm$ 0.11 & - & 0.184 (*)\\
  9655005 & 0.71508 & 0.040 $\pm$ 0.007 & 1.42971 & 0.013 $\pm$ 0.001 & 8086 $\pm$ 78&3.732 $\pm$ 0.162 & 1.451 $\pm$ 0.049 & 2.71 $\pm$ 0.07 & 98 $\pm$ 1 &$<$ 120 & 1.87 $\pm$ 0.12 & 1.43 $\pm$ 0.106 & 1.505 $\pm$ 0.116\\
  9711038 & 0.67416 & 0.012 $\pm$ 0.001 & 1.34825 & 0.003 $\pm$ 0.001 & 7775 $\pm$ 294&3.736 $\pm$ 0.255& 1.154 $\pm$ 0.047 & 2.08 $\pm$ 0.16 & 71 $\pm$ 1 &$<$ 120& 0.79 $\pm$ 0.12 & - & 3.060 (*)\\
  9760777 & 1.69826 & 0.031 $\pm$ 0.001 & 3.41022 & 0.004 $\pm$ 0.001 & 8107 $\pm$ 283&3.799 $\pm$ 0.224 & 1.456 $\pm$ 0.047 & 2.71 $\pm$ 0.20 & 233 $\pm$ 3&$<$ 120 & 1.65 $\pm$ 0.12 & 0.572 $\pm$ 0.168 & 0.826 $\pm$ 0.109\\
  10068389 & 1.01706 & 0.051 $\pm$ 0.001 & 2.03422 & 0.006 $\pm$ 0.002 & 8647 $\pm$ 308&3.787 $\pm$ 0.275 & 1.595 $\pm$ 0.043 & 2.80 $\pm$ 0.20 & 144 $\pm$ 2&$<$ 120 & 2.18 $\pm$ 0.12 & 5.451 $\pm$  0.103 & 0.554 $\pm$ 0.116\\
  10153555 & 0.94086 & 0.009 $\pm$ 0.001 & 1.77012 & 0.005 $\pm$ 0.001 & 7800 $\pm$ 270&4.042 $\pm$ 0.160 & 1.253 $\pm$ 0.045 & 2.32 $\pm$ 0.17 & 111 $\pm$ 1 &- & 0.76 $\pm$ 0.12 & - & 0.829 (*)\\
  10156332 & 1.21457 & 0.017 $\pm$ 0.001 & 2.42931 & 0.007 $\pm$ 0.001 & 8501 $\pm$ 321&4.211 $\pm$ 0.130 & 1.263 $\pm$ 0.049 & 1.98 $\pm$ 0.15 & 122 $\pm$ 3 &$<$ 120 & 0.89 $\pm$ 0.12 & - & 1.342 (*)\\
  10215038 & 1.40541 & 0.020 $\pm$ 0.001 & 2.81120 & 0.006 $\pm$ 0.001 & 9160 $\pm$ 78&3.756 $\pm$ 0.152 & 1.486 $\pm$ 0.049 & 2.20 $\pm$ 0.05 & 157 $\pm$ 1& $<$ 120 & 1.07 $\pm$ 0.11 & - & 3.974 (*)\\
  10216016 & 0.93120 & 0.029 $\pm$ 0.001 & 1.86244 & 0.006 $\pm$ 0.001 & 7823 $\pm$ 274&4.006 $\pm$ 0.170 & 1.306 $\pm$ 0.048 & 2.45 $\pm$ 0.18 & 116 $\pm$ 2 &-& 1.44 $\pm$ 0.12 & 1.001 $\pm$ 0.109 & 0.679 $\pm$ 0.107\\
  10338005 & 2.09676 & 0.026 $\pm$ 0.001 & 4.19398 & 0.006 $\pm$ 0.001 & 10000 $\pm$ 357&3.658 $\pm$ 0.250 & 1.989 $\pm$ 0.102 & 3.30 $\pm$ 0.25 & 350 $\pm$ 5 &- & 1.84 $\pm$ 0.13 & 0.409 $\pm$ 0.111 & 0.573 $\pm$ 0.104\\
  10354997 & 1.85626 & 0.006 $\pm$ 0.001 & 3.71121 & 0.003 $\pm$ 0.001 & 8698 $\pm$ 326&4.001 $\pm$ 0.171 & 1.223 $\pm$ 0.043 & 1.80 $\pm$ 0.14 & 169 $\pm$ 3 &$<$ 120 & 0.48 $\pm$ 0.12 & 0.531 $\pm$  0.171 & 0.635 $\pm$ 0.107\\
  10394172 & 0.75101 & 0.016 $\pm$ 0.003 & 1.48950 & 0.005 $\pm$ 0.001 & 9160 $\pm$ 346&4.106 $\pm$ 0.164 & 1.479 $\pm$ 0.058 & 2.18 $\pm$ 0.17 & 83 $\pm$ 1 &- & 0.95 $\pm$ 0.13 & - & 1.011 (*)\\
  10405887 & 1.05742 & 0.022 $\pm$ 0.001 & 2.11529 & 0.008 $\pm$ 0.001 & 9282 $\pm$ 362&3.877 $\pm$ 0.250 & 1.692 $\pm$ 0.053 & 2.72 $\pm$ 0.22 & 145 $\pm$ 2 &$<$ 120 & 1.39 $\pm$ 0.12 & 0.920 $\pm$ 0.107 & 0.820 $\pm$ 0.108 \\
  10467815 & 1.42021 & 0.028 $\pm$ 0.001 &2.84029& 0.004 $\pm$ 0.001 & 8932 $\pm$ 358&3.822 $\pm$ 0.257 & 1.354 $\pm$ 0.044 & 1.99 $\pm$ 0.16 & 143 $\pm$ 2 &$<$ 120 & 1.15 $\pm$ 0.12 & 0.674 $\pm$ 0.109 & 0.608 $\pm$ 0.106\\
  10529091 & 0.71393 & 0.037 $\pm$ 0.001 & 1.42817 & 0.005 $\pm$ 0.001 & 8004 $\pm$ 280&3.720 $\pm$ 0.275 & 1.466 $\pm$ 0.046 & 2.82 $\pm$ 0.20 & 102 $\pm$ 1 &- & 1.87 $\pm$ 0.12 & 1.389 $\pm$  0.103& 2.553 $\pm$ 0.126\\
  10533233 & 1.13379 & 0.027 $\pm$ 0.002 & 2.26772 & 0.005 $\pm$ 0.002 & 9710 $\pm$ 374&3.720 $\pm$ 0.275 & 1.797 $\pm$ 0.062 & 2.80 $\pm$ 0.22 & 161 $\pm$ 2 &$<$ 120 & 1.59 $\pm$ 0.13 & 0.858 $\pm$  0.105 & 0.714 $\pm$ 0.107\\
  10548172 & 2.06429 & 0.014 $\pm$ 0.001 & 4.12846 & 0.008 $\pm$ 0.001 & 8071 $\pm$ 294&4.103 $\pm$ 0.133 & 1.307 $\pm$ 0.049 & 2.31 $\pm$ 0.17 & 241 $\pm$ 4 &- & 0.94 $\pm$ 0.12 & - & 3.537 (*)\\
  10612854 & 1.52043 & 0.007 $\pm$ 0.001 & 3.04376 & 0.002 $\pm$ 0.001 & 8585 $\pm$ 82&3.766 $\pm$ 0.165& 1.460 $\pm$ 0.051 & 2.43 $\pm$ 0.06 & 187 $\pm$ 1 &$<$ 120 & 0.70 $\pm$ 0.11 & - & 1.761 (*)\\
  10810140 & 1.48415 & 0.009 $\pm$ 0.001 & 2.96864 & 0.003 $\pm$ 0.001 & 8265 $\pm$ 82&3.735 $\pm$ 0.210& 1.146 $\pm$ 0.044 & 1.83 $\pm$ 0.05 & 137 $\pm$ 1 &$<$ 120 & 0.60 $\pm$ 0.11 & 0.654 $\pm$ 0.105 & 0.662 $\pm$ 0.108\\
  10816270 & 1.14221 & 0.081 $\pm$ 0.001 & 2.29810 & 0.013 $\pm$ 0.001 & 9345 $\pm$ 376&4.027 $\pm$ 0.195& 1.556 $\pm$ 0.048 & 2.29 $\pm$ 0.19 & 132 $\pm$ 2 &- & 2.25 $\pm$ 0.12 & 0.858 $\pm$ 0.104 & 0.974 $\pm$ 0.113\\
  10816278 & 1.06760 & 0.009 $\pm$ 0.001 & 2.13659 & 0.009 $\pm$ 0.001 & 8060 $\pm$ 296&3.843 $\pm$ 0.215 & 1.094 $\pm$ 0.045 & 1.81 $\pm$ 0.14 & 98 $\pm$ 2 &- & 0.59 $\pm$ 0.12 & - & 1.306 (*)\\
  10817581 & 0.99167 & 0.005 $\pm$ 0.001 & 1.97581 & 0.003 $\pm$ 0.001 & 8524 $\pm$ 306&3.793 $\pm$ 0.252 & 1.462 $\pm$ 0.049 & 2.47 $\pm$ 0.18 & 124 $\pm$ 2 &- & 0.60 $\pm$ 0.10 & 0.920 $\pm$ 0.118 & 0.700 $\pm$  0.108\\
  10879812 & 1.42972 & 0.011 $\pm$ 0.001 & 2.85941 & 0.001 $\pm$ 0.001 & 8250 $\pm$ 283&4.089 $\pm$ 0.145 & 1.271 $\pm$ 0.042 & 2.12 $\pm$ 0.15 & 153 $\pm$ 3&- & 0.77 $\pm$ 0.12 & 0.695 $\pm$  0.193 & 0.938 $\pm$ 0.112\\
  10974769 & 1.76150 & 0.029 $\pm$ 0.001 & 3.52144 & 0.005 $\pm$ 0.001 & 8681 $\pm$ 326&3.973 $\pm$ 0.185 & 1.312 $\pm$ 0.046 & 2.00 $\pm$ 0.15 & 179 $\pm$ 3 &$<$ 120 & 1.18 $\pm$ 0.12 & 0.572 $\pm$ 0.111 & 0.930 $\pm$ 0.111\\
  10990092 & 0.84952 & 0.007 $\pm$ 0.001 & 1.70125 & 0.006 $\pm$ 0.001 & 8059 $\pm$ 281&3.731 $\pm$ 0.275 & 1.236 $\pm$ 0.047 & 2.13 $\pm$ 0.15 & 92 $\pm$ 1 &- & 0.62 $\pm$ 0.12 & - & 1.174 (*)\\
  11091033 & 1.32990 & 0.013 $\pm$ 0.001 & 2.65951 & 0.002 $\pm$ 0.001 & 8856$\pm$ 338&4.043 $\pm$ 0.170 & 1.42 $\pm$ 0.046 & 2.18 $\pm$ 0.17 & 147 $\pm$ 2 &- & 0.86 $\pm$ 0.11 & 0.736 $\pm$ 0.108 & 0.698 $\pm$ 0.108\\
  11288072 & 0.88564 & 0.118 $\pm$ 0.001 & 1.77149 & 0.018 $\pm$ 0.001 & 8111 $\pm$ 285&3.724 $\pm$ 0.270 & 1.514 $\pm$ 0.047 & 2.90 $\pm$ 0.21 & 130 $\pm$ 2  &124 $\pm$ 79& 3.44 $\pm$ 0.12 & 1.124 $\pm$  0.103 & 11.011 $\pm$ 0.217\\
  11296464 & 1.55118 & 0.015 $\pm$ 0.005 & 3.09483 & 0.005 $\pm$ 0.001 & 8363 $\pm$ 80&3.725 $\pm$ 0.160 & 1.266 $\pm$ 0.045 & 2.05 $\pm$ 0.05 & 161 $\pm$ 1 &149 $\pm$ 51& 0.86 $\pm$ 0.11 & 0.633 $\pm$  0.121 & 0.873 $\pm$ 0.101\\
  11303065 & 0.89503 & 0.023 $\pm$ 0.001 & 1.79012 & 0.014 $\pm$ 0.001 & 7679 $\pm$ 270&3.818 $\pm$ 0.225 & 0.959 $\pm$ 0.043 & 1.71 $\pm$ 0.13 & 77 $\pm$ 1 &$<$ 120& 0.90 $\pm$ 0.11 & - & 3.985 (*)\\
  11357670 & 1.35987 & 0.063 $\pm$ 0.001 & 2.72008 & 0.026 $\pm$ 0.001 & 8310 $\pm$ 79&3.908 $\pm$ 0.109 & 1.213 $\pm$ 0.044 & 1.95 $\pm$ 0.05 & 134 $\pm$ 1 &$<$ 120 & 1.69 $\pm$ 0.11 & 0.715 $\pm$ 0.102 & 0.850 $\pm$ 0.110\\
  11442175 & 1.37440 & 0.013 $\pm$ 0.001 & 2.74830 & 0.007 $\pm$ 0.001 & 8421 $\pm$ 302&3.762 $\pm$ 0.280 & 1.193 $\pm$ 0.044 & 1.86 $\pm$ 0.14 & 129 $\pm$ 2 &- & 0.73 $\pm$ 0.12 & 0.674 $\pm$ 0.110 & 0.644 $\pm$ 0.107\\
  11443271 & 1.27051 & 0.021 $\pm$ 0.001 & 2.54053 & 0.009 $\pm$ 0.001 & 8331 $\pm$ 277&3.777 $\pm$ 0.255 & 1.784 $\pm$ 0.042 & 3.75 $\pm$ 0.25 & 241 $\pm$ 2 &180 & 1.88 $\pm$ 0.13 & 0.797 $\pm$  0.102 & 9.171 $\pm$ 0.176\\
  12072819 & 1.27417 & 0.032 $\pm$ 0.001 & 2.54851 & 0.005 $\pm$ 0.001 & 8145 $\pm$ 78&3.954 $\pm$ 0.105 & 1.151 $\pm$ 0.045 & 1.89 $\pm$ 0.05 & 122 $\pm$ 1 &$<$ 120 & 1.17 $\pm$ 0.11 & 0.695 $\pm$  0.111 & 0.780 $\pm$ 0.108\\
  12117276 & 1.50762 & 0.055 $\pm$ 0.001 & 3.01504 & 0.056 $\pm$ 0.001 & 7679 $\pm$ 76&3.979 $\pm$ 0.135 & 1.358 $\pm$ 0.048 & 2.70$\pm$ 0.07 & 206 $\pm$ 2 &129 & 2.19 $\pm$ 0.11 & 0.654 $\pm$  0.139 &  0.748 $\pm$ 0.108\\
  12155426 & 0.88528 & 0.009 $\pm$ 0.001 & 1.71389 & 0.005 $\pm$ 0.001 & 7882 $\pm$ 288&3.712 $\pm$ 0.278 & 1.451 $\pm$ 0.047 & 2.86 $\pm$ 0.21 & 128 $\pm$ 2 &- & 0.94 $\pm$ 0.10 & - & 0.651 (*)\\
  12306265 & 1.12762 & 0.006 $\pm$ 0.001 & 2.25611 & 0.001 $\pm$ 0.001 & 8074 $\pm$ 294&3.780 $\pm$ 0.250 & 1.331 $\pm$ 0.042 & 2.37 $\pm$ 0.18 & 135 $\pm$ 2 &- & 0.63 $\pm$ 0.10 & - & 6.083 (*)\\
  12602335 & 1.47120 & 0.008 $\pm$ 0.001 & 2.94340 & 0.007 $\pm$ 0.001 & 8151 $\pm$ 306&3.905 $\pm$ 0.190 & 1.382 $\pm$ 0.044 & 2.47 $\pm$ 0.19 & 184 $\pm$ 2 &$<$ 120 & 0.76 $\pm$ 0.12 & 0.613 $\pm$  0.110 & 0.783 $\pm$ 0.109\\
\hline
\end{tabular}
\end{table}
\end{landscape}

\begin{landscape}
\begin{table}
\caption{List of Am/Fm stars with ``hump and spike'' features in the frequency spectra.  We list the rotation frequency ($\rm f_{rot}$), rotational velocity ($\rm V_{rot}$), rotation frequency amplitude ($\rm A_{rot}$), frequency of the first harmonic of the rotation frequency ($\rm f_{har}$) and its amplitude ($\rm A_{har}$) as derived from their frequency spectra, the effective temperature ($\rm T_{eff}$) and surface gravity (log\,g) as reported by \citet{2017ApJS..229...30M}, the luminosity (log(L/$\rm L_{\odot}$)) as derived from Gaia parallaxes \citep{2018yCat.1345....0G}, the stellar radius (R) as estimated from the effective temperature and luminosity, the projected rotational velocity ($\nu $~sin~$i$) as found in the literature \citep{2016A&A...594A..39F} in general; \citet{10.1093/mnras/stv528} for KIC\,9117875 and \citet{2017MNRAS.470.2870N} for KIC\,9349245, the spot radius ($\rm R_{spot}$) as estimated from the photometric amplitude, the rotation period ($\rm P_{ACF}$) and decay-time scale of the starspots ($\rm \tau_{DT}$) as estimated from the autocorrelation functions of their light-curves. (*) marks doubtful measurements.}
\fontsize{7.5}{9.5}\selectfont
\begin{tabular}{lccccccccccccr}
\hline\hline\\
 KIC &$\rm f_{rot}$ &$\rm A_{rot}$ &$\rm f_{har}$ &$\rm A_{har}$ &$\rm T_{eff}$&log\,g&log(L/$\rm L_{\odot}$)&R&$\rm V_{rot}$&$\nu $~sin~$i$&$\rm R_{spot}$ &$\rm P_{ACF}$&$\rm \tau_{DT}$ \\\\
 &$\rm (d^{-1})$&$\rm (mmag)$&$\rm (d^{-1})$&$\rm (mmag)$&(K)&(dex)&(dex)&($\rm R_{\odot}$)&$\rm (km~s^{-1})$&$\rm (km~s^{-1})$&($\rm R_{E}$)&(d)&(d)\\
\hline
  3238787 & 1.72468 & 0.019 $\pm$ 0.004 & - & - & 6760 $\pm$ 81&3.934 $\pm$ 0.158 & 1.305 $\pm$ 0.048 & 3.28 $\pm$ 0.10 & 286 $\pm$ 2 & $<$ 120 & 1.56 $\pm$ 0.12 & - & 1.852 (*)\\
  3240406 & 1.06809 & 0.005 $\pm$ 0.001 & 2.14095 & 0.004 $\pm$ 0.001 & 7936 $\pm$ 80&4.029 $\pm$ 0.110 & 1.128 $\pm$ 0.043 & 1.94 $\pm$ 0.05 & 105 $\pm$ 1 & $<$ 120 & 0.47 $\pm$ 0.11 & - & 1.872 (*)\\
  3956495 & 1.07119 & 0.031 $\pm$ 0.001 & 2.13973 & 0.019 $\pm$ 0.001 & 7021 $\pm$ 237&3.827 $\pm$ 0.18& 1.337 $\pm$ 0.131 & 3.15 $\pm$ 0.26 & 171 $\pm$ 2& $<$ 120 & 1.92 $\pm$ 0.13 & 0.899 $\pm$  0.114 &  0.711 $\pm$ 0.117\\
  4770092 & 0.73973 & 0.023 $\pm$ 0.002 & 1.47906 & 0.016 $\pm$ 0.002 & 6777 $\pm$ 205&3.669 $\pm$ 0.180 & 1.183 $\pm$ 0.044 & 2.84 $\pm$ 0.18 & 107 $\pm$ 1 & $<$ 120 & 1.49 $\pm$ 0.13 & -  & 2.241 (*)\\
  4839729 & 1.34295 & 0.009 $\pm$ 0.001 & 2.68266 & 0.004 $\pm$ 0.001 & 6815 $\pm$76&4.093 $\pm$ 0.125 & 0.663 $\pm$ 0.042 & 1.54 $\pm$ 0.06 & 105 $\pm$ 1 & $<$ 120 & 0.50 $\pm$ 0.11 & 0.797 $\pm$  0.111 & 1.358 $\pm$ 0.128\\
  5121064 & 0.74769 & 0.017 $\pm$ 0.002 & 1.49106 & 0.009 $\pm$ 0.002 & 6666 $\pm$ 79&4.114 $\pm$ 0.124 & 1.113 $\pm$ 0.046 & 2.70 $\pm$ 0.09 & 102 $\pm$ 1& $<$ 120 & 1.22 $\pm$ 0.11 & 1.451 $\pm$  0.116 & 11.674 $\pm$ 0.197\\
  6039039 & 0.89373 & 0.016 $\pm$ 0.001 & - & -  & 7786 $\pm$ 79&3.956 $\pm$ 0.110 & 1.250 $\pm$ 0.042 & 2.32 $\pm$ 0.06 & 105 $\pm$ 1 & $<$ 120 & 1.01 $\pm$ 0.11 & - & 11.807 (*)\\
  6209721 & 0.65867 & 0.010 $\pm$ 0.001 & 1.32945 & 0.005 $\pm$ 0.001 & 6775 $\pm$ 79&4.055 $\pm$ 0.138 & 1.190 $\pm$ 0.046 & 2.86 $\pm$ 0.09 & 95 $\pm$ 1& $<$ 120 & 0.99 $\pm$ 0.11 & 1.349 $\pm$  0.121 & 0.594 $\pm$ 0.118\\
  6804592 & 0.73354 & 0.005 $\pm$ 0.001 & 1.46787 & 0.002 $\pm$ 0.001 & 8557 $\pm$ 316&4.094 $\pm$ 0.146 & 1.391 $\pm$ 0.045 & 2.26 $\pm$ 0.17 & 84 $\pm$ 1 & $<$ 120 & 0.55 $\pm$ 0.12 & - & 0.902 (*)\\
  6960377 & 1.03100 & 0.022 $\pm$ 0.001 & 2.06143 & 0.006 $\pm$ 0.001 & 7780 $\pm$ 286&3.785 $\pm$ 0.230 & 1.239 $\pm$ 0.046 & 2.30 $\pm$ 0.17 & 120 $\pm$ 2 & $<$ 120 & 1.18 $\pm$ 0.12 &- & 2.757 (*)\\
  7287683 & 0.94411 & 0.008 $\pm$ 0.001 & 1.88856 & 0.006 $\pm$ 0.001 & 7651 $\pm$ 268&4.075 $\pm$ 0.150 & 1.120 $\pm$ 0.046 & 2.07 $\pm$ 0.15 & 99 $\pm$ 1 & $<$ 120 & 0.64 $\pm$ 0.12 & 1.042 $\pm$ 0.113 & 0.821 $\pm$ 0.119\\
  7287786 & 0.75129 & 0.015 $\pm$ 0.001 & 1.50261 & 0.008 $\pm$ 0.001 & 7864 $\pm$ 84&3.841 $\pm$ 0.147 & 0.984 $\pm$ 0.048 & 1.68 $\pm$ 0.05 & 64 $\pm$ 1 & $<$ 120 & 0.71$\pm$ 0.11 & - & 3.685 (*)\\
  7466060 & 1.09884 & 0.065 $\pm$ 0.005 & -& - & 6844 $\pm$ 242&4.240 $\pm$ 0.121 & 1.253 $\pm$ 0.051 & 3.01 $\pm$ 0.22 & 168 $\pm$ 2 & $<$ 120 & 2.65 $\pm$ 0.15 & 1.124 $\pm$  0.177 & 5.485 $\pm$ 0.159\\
  8429756 & 0.36458 & 0.031 $\pm$ 0.003 & 0.71879 & 0.012 $\pm$ 0.003 & 7505 $\pm$ 78&4.342 $\pm$ 0.067 & 1.013 $\pm$ 0.043 & 1.90 $\pm$ 0.06 & 35 $\pm$ 1 & - & 1.16 $\pm$ 0.11 & 2.899 $\pm$  0.117 & 8.402 $\pm$ 0.159\\
  8519992 & 0.45862 & 0.024 $\pm$ 0.003 & 0.92039 & 0.012 $\pm$ 0.003 & 6788 $\pm$ 271&3.789 $\pm$ 0.280 & 1.313 $\pm$ 0.047 & 3.28 $\pm$ 0.27 & 76 $\pm$ 1 & $<$ 120 & 1.75 $\pm$ 0.19 & 2.656 $\pm$  0.116 &10.147 $\pm$ 0.224\\
  9273647 & 0.92991 & 0.011 $\pm$ 0.001 & 1.86037 & 0.006 $\pm$ 0.001 & 7046 $\pm$ 252&3.905 $\pm$ 0.250 & 1.375 $\pm$ 0.042 & 3.27 $\pm$ 0.24 & 154 $\pm$ 2 & $<$ 120 & 1.18 $\pm$ 0.10 &1.103 $\pm$  0.115 & 4.251 $\pm$ 0.152\\
  9906894 & 1.68073 & 0.036 $\pm$ 0.002 & 3.36628 & 0.044 $\pm$ 0.002 & 7615 $\pm$ 79&4.045 $\pm$ 0.120 & 1.409 $\pm$ 0.047 & 2.91 $\pm$ 0.08 & 248 $\pm$ 2& $<$ 120 & 1.91 $\pm$ 0.11 & 0.552 $\pm$  0.117 & 0.782 $\pm$ 0.119\\
  10015534 & 1.19377 & 0.017 $\pm$ 0.002 & 2.39360 & 0.005 $\pm$ 0.002 & 8028 $\pm$ 294&4.041 $\pm$ 0.158 & 1.057 $\pm$ 0.044 & 1.75 $\pm$ 0.13 & 106 $\pm$ 2 &$<$ 120 & 0.79 $\pm$ 0.12 & 1.022 $\pm$  0.115 & 5.242 $\pm$ 0.186\\
  10129532 & 0.75688 & 0.006 $\pm$ 0.001 & 1.51573 & 0.002 $\pm$ 0.001 & 7929 $\pm$ 279&3.831 $\pm$ 0.225 & 1.319 $\pm$ 0.055 & 2.42 $\pm$ 0.18 & 93 $\pm$ 1 &$<$ 120 & 0.65 $\pm$ 0.12 & - & 1.848 (*)\\
  11027806 & 1.11659 & 0.009 $\pm$ 0.003 & 2.22648 & 0.003 $\pm$ 0.003 & 9080 $\pm$ 340&3.405 $\pm$ 0.457 & 2.318 $\pm$ 0.087 & 5.83 $\pm$ 0.45 & 330 $\pm$ 3 &273 $\pm$ 88 & 1.91 $\pm$ 0.19 & - & 1.196 (*)\\
  11072219 & 0.87721 & 0.026 $\pm$ 0.002 & 1.64625 & 0.008 $\pm$ 0.002 & 6797 $\pm$ 193&3.807 $\pm$ 0.190 & 0.675 $\pm$ 0.042 & 1.57 $\pm$ 0.10 & 70 $\pm$ 1 & $<$ 120 & 0.87 $\pm$ 0.12 & 1.369 $\pm$  0.116 & 1.133 $\pm$  0.122\\
  9349245 & 1.10719 & 0.005 $\pm$ 0.001 & 2.21412 & 0.004 $\pm$ 0.001 & 7834 $\pm$ 278&3.723 $\pm$ 0.294 & 1.012 $\pm$ 0.041 & 1.74 $\pm$ 0.13 & 98 $\pm$ 2 & 82 $\pm$ 2 & 0.42 $\pm$ 0.10 & 0.879 $\pm$  0.121 & 0.683 $\pm$ 0.118\\
  3459226 & 0.92658 & 0.041 $\pm$ 0.003 & 1.85341 & 0.009 $\pm$ 0.003 & 7280 $\pm$ 82&3.622 $\pm$ 0.165 & 0.886 $\pm$ 0.046 & 1.75 $\pm$ 0.06 & 82 $\pm$ 1 & $<$ 120 & 1.22 $\pm$ 0.11 & 1.042 $\pm$  0.116 & 0.860 $\pm$  0.121\\
  9272082 & 1.03583 & 0.001 $\pm$ 0.001 & 2.08237 & 0.001 $\pm$ 0.001 & 9077 $\pm$ 333&4.147 $\pm$ 0.154 & 1.852 $\pm$ 0.043 & 3.42 $\pm$ 0.25 & 179 $\pm$ 2 & $<$ 120 & 0.37 $\pm$ 0.10 & - & 1.693 (*)\\
  5302167 & 2.27153 & 0.007 $\pm$ 0.001 & 4.56527 & 0.004 $\pm$ 0.001 & 8402 $\pm$ 303&3.930 $\pm$ 0.190&1.491 $\pm$ 0.083 & 2.63 $\pm$ 0.71 & 302 $\pm$ 14 & $<$ 120 & 0.76 $\pm$ 0.18 & - & 1.698 (*)\\
  6116612 & 0.59486 & 0.024 $\pm$ 0.001 & 1.18968 & 0.011 $\pm$ 0.001 & 7012  $\pm$ 264&4.088 $\pm$ 0.205&0.797 $\pm$ 0.043 & 1.70 $\pm$ 0.44 & 51 $\pm$ 2 & $<$ 120 & 0.91 $\pm$ 0.15 & - & 2.100 (*)\\
  8110941 & 0.78137 & 0.046 $\pm$ 0.010 & 1.55609 & 0.010 $\pm$ 0.010 & 7163 $\pm$ 257&4.143 $\pm$ 0.165 &0.840 $\pm$ 0.047& 1.71 $\pm$ 0.46 & 67 $\pm$ 2& $<$ 120 & 1.27 $\pm$ 0.26 & 1.450 $\pm$ 0.188 & 5.467 $\pm$ 0.154 \\
  8715392 & 0.79754 & 0.016 $\pm$ 0.001 & 1.59484 & 0.003 $\pm$ 0.001 & 6889 $\pm$ 239&3.937 $\pm$ 0.295&0.831 $\pm$ 0.056 & 1.83 $\pm$ 0.60 & 74 $\pm$ 2 &$<$ 120 & 0.80 $\pm$ 0.17 & -  & 4.010 (*)\\
  9426071 & 0.38520 & 0.032 $\pm$ 0.004 & 0.76571 & 0.015 $\pm$ 0.004 & 6706 $\pm$ 213&4.226 $\pm$ 0.157 &0.564 $\pm$ 0.029 & 1.42 $\pm$ 0.35 & 28 $\pm$ 1 & $<$ 120 & 0.88 $\pm$ 0.18 & - & 0.808 (*)\\
  11551962 & 1.70180 & 0.011 $\pm$ 0.002 & 3.40584 & 0.004 $\pm$ 0.002 & 6728 $\pm$ 225&4.161 $\pm$ 0.180& 0.065 $\pm$ 0.037 & 1.56 $\pm$ 0.43 & 134 $\pm$ 6 &$<$ 120 & 0.59 $\pm$ 0.17 & 0.531  $\pm$ 0.128 &  0.772 $\pm$ 0.118\\
  9238276 & 0.37650 & 0.270 $\pm$ 0.022 & 0.75927 & 0.099 $\pm$ 0.022 & 6038 $\pm$ 172&4.524 $\pm$ 0.126 &-0.015 $\pm$ 0.060 & 0.90 $\pm$ 0.17 & 17 $\pm$ 1 & $<$ 122 & 1.61 $\pm$ 0.19 & 2.983 $\pm$ 0.110 & 7.202 $\pm$ 0.166\\
  4364400 & 0.36372 &0.028 $\pm$ 0.003 & 0.72262& 0.010 $\pm$ 0.003 &5802 $\pm$ 165&4.327 $\pm$ 0.165 &0.129 $\pm$ 0.062& 1.15 $\pm$ 0.27 & 21 $\pm$ 2 &-& 0.66 $\pm$ 0.13 & - &  1.616 (*)\\
  5726737 & 0.48377 &0.027 $\pm$  0.002& 0.95797 &0.011 $\pm$ 0.002& 6015 $\pm$ 211&4.196 $\pm$ 0.200& 0.344 $\pm$ 0.020 & 1.37 $\pm$ 0.38 & 34 $\pm$ 4 &-& 0.78 $\pm$ 0.15 & 1.921 $\pm$  0.128 & 0.528 $\pm$ 0.121\\
  6042168 & 0.95803 & 0.024 $\pm$ 0.002 & 1.93109 & 0.009 $\pm$ 0.002 & 6796 $\pm$ 195&3.430 $\pm$ 0.215 & 0.964 $\pm$ 0.046 & 2.19 $\pm$ 0.14 & 106 $\pm$ 1&- & 1.17 $\pm$ 0.12 & 0.920 $\pm$ 0.112 & 0.814 $\pm$ 0.119\\
  7116117 & 0.94531 & 0.008 $\pm$ 0.001 & 1.88832 & 0.001 $\pm$ 0.001 & 7518 $\pm$ 276&4.023 $\pm$ 0.160 & 1.097 $\pm$ 0.043 & 2.09 $\pm$ 0.16 & 100 $\pm$ 1&- & 0.65 $\pm$ 0.10 & 1.022 $\pm$  0.116 & 0.694 $\pm$ 0.119\\
  8655763 & 1.45803 & 0.021 $\pm$ 0.002 & 2.91586 & 0.003 $\pm$ 0.002 & 8235 $\pm$ 310&3.665 $\pm$ 0.280 & 1.488 $\pm$ 0.052 & 2.73 $\pm$ 0.21 & 201 $\pm$ 3 &- & 1.37 $\pm$ 0.13 & - & 2.501 (*)\\
  9655438 & 2.12307 & 0.126 $\pm$ 0.002 & 4.24670 & 0.022 $\pm$ 0.002 & 7118 $\pm$ 253&4.074 $\pm$ 0.184 & 1.305 $\pm$ 0.051 & 2.96 $\pm$ 0.22 & 318 $\pm$ 5 &- & 3.63 $\pm$ 0.13 & 0.467 $\pm$ 0.116 & 7.489 $\pm$ 0.208\\
  10357965 & 0.63053 & 0.051 $\pm$ 0.002 & 1.26102 & 0.026 $\pm$ 0.002 & 7210 $\pm$ 275&4.239 $\pm$ 0.152 & 1.368 $\pm$ 0.052 & 3.10 $\pm$ 0.24 & 99 $\pm$ 1 &- & 2.42 $\pm$ 0.14 & - & 1.540 (*)\\
  9117875& 0.71566 & 0.046 $\pm$ 0.001& 1.43098 & 0.021 $\pm$ 0.001& 7300 $\pm$ 226&3.931 $\pm$ 0.260& 1.010 $\pm$ 0.050 & 1.98 $\pm$ 0.64 &72 $\pm$ 2&61 $\pm$ 3& 1.47 $\pm$ 0.16 & 1.451 $\pm$  0.134 & 4.852 $\pm$ 0.161\\
\hline\end{tabular}
\label{table_Am}
\end{table}
\end{landscape}



%
%
%
 \bibliographystyle{mnras}
 \bibliography{ref} 
%
%
%
%

%
%
%
%
%

 \bsp	
 \label{lastpage}
 \end{document}